\documentclass[]{elsarticle} 
\usepackage{graphicx}
\usepackage{xcolor}
\usepackage{soul}
\usepackage{amssymb}   
\usepackage{amsmath}
\usepackage[caption=false]{subfig}
\usepackage[a4paper, inner=2.50cm, outer=2.5cm, top=2.5cm,bottom=2.5cm, binding offset=0 cm]{geometry}
\begin{document}

\title{Analytic and numeric computation of edge states and
  conductivity of a Kane-Mele nanoribbon}

\author[rvt]{Priyanka Sinha}
\ead{sinhapriyanka2016@iitg.ernet.in}

\author[rvt]{Sudin Ganguly}
\ead{sudin@iitg.ernet.in}

\author[rvt]{Saurabh Basu}
\ead{saurabh@iitg.ernet.in}

\address{$^1$Indian Institute of Technology Guwahati,  Guwahati,
  Assam-781039, India}

\begin{abstract}
We compute analytic expressions for the edge states in a zigzag
Kane-Mele nanoribbon (KMNR) by solving the eigenvalue equations in
presence of intrinsic and Rashba spin-orbit couplings. Owing to the
P-T symmetry of the Hamiltonian the edge states are protected by
topological invariance and hence are found to be robust. We have done a systematic study for each of the above cases, for example, a pristine graphene, graphene with an intrinsic spin-orbit coupling, graphene with a Rashba spin-orbit coupling, a Kane-Mele nanoribbon and supported our results on the
robustness of the edge states by analytic computation of the
electronic probability amplitudes, the local density of states (LDOS),
band structures and the conductance spectra.
\end{abstract}
\begin{keyword}
  Kane-Mele nanoribbon\sep Band structure\sep Electronic wavefunction\sep LDOS\sep Charge conductance
\end{keyword}

\maketitle
\section{\label{sec1}Introduction}
The successful fabrication of graphene~\cite{novo} has generated
intense research activities to study the electronic properties of this
novel two dimensional (2D) electronic system. Graphene has a honeycomb
lattice structure due to the $sp^2$ hybridization of carbon atoms and
the $\pi$-electrons can hop between nearest neighbors. The valence and
conduction bands of graphene touch each other at two nonequivalent
Dirac points, $K$ and ${{K}^{\prime}}$, which have opposite
chiralities and form a time-reversed pair. The band structure around
those points has the Dirac form, ${{E}_{{\vec{k}}}}=\hbar v|\vec{k}|$
, where $v$ $(\simeq {{10}^{6}}$ ms$^{-1}$) is the Fermi velocity. The
Dirac nature of the electrons~\cite{wallace} is responsible for many
interesting properties of graphene~\cite{neto}, such as unconventional
quantum Hall effect~\cite{novo,zhang,vp}, half metallicity~\cite
{jun,lin}, Klein tunneling through a barrier~\cite{kat}, high carrier
mobility~\cite{du,bolotin} and many more. Owing to these features,
graphene is recognized as one of the promising materials for realizing
next-generation electronic devices.

The existence of edge states in a graphene sheet is one of the
interesting features in condensed matter physics. The properties of
the edge states are different than the bulk states and play important
roles in transport.  When the valence and conduction bands are
separated by an energy gap, electrons can not flow through the
bulk. However, this does not guarantee that the system is a simple
insulator, since conduction may still be allowed via edge modes. These
new type of insulators are different from trivial insulators due to
their unique gapless edge states protected by the time-reversal
symmetry and they are attributed the name {\it{topological
    insulators}}. Thus topological insulators (TIs) represent a new
quantum state. The phenomena associated with the TIs are the well
known quantum Hall effect and the quantum spin Hall effect, where it
has been found that the gapless chiral or helical edge states are
robust channels with quantized conductance accompanied by
non-vanishing values at zero
bias~\cite{halperin,laughlin,kane-mele1,kane-mele2,bernevig,bhz,wu,onada}.

Kane and Mele~\cite{kane-mele1,kane-mele2} predicted that a quantum
spin Hall (QSH) state can be observed in presence of next-nearest
neighbour intrinsic spin-orbit coupling (SOC), which triggered an
enormous study on topologically non-trivial electronic
materials~\cite{bernevig,moore,hasan,qi}. Unfortunately, the QSH phase
in pristine graphene is still not observed experimentally owing to its
vanishingly small intrinsic spin-orbit coupling strength (typically
$\sim$ 0.01-0.05 meV)~\cite{min,yao}, whereas in strong SOC materials,
such as CdTe/HgTe quantum wells, the QSH phase has been
observed~\cite{konig}. However, there are various ways to induce SOC
in graphene experimentally, such as doping by adatoms~\cite{weeks},
using the proximity to a three-dimensional topological insulator,
such as Bi$_2$Se$_3$~\cite{kou,j-zhang}, by functionalization with
methyl~\cite{zollner} etc.

Two kinds of SOC can be present in graphene, the intrinsic and the
Rashba SOC. Recent observations showed that the strength of the Rashba
SOC can also be enhanced up to 100 meV from Gold (Au) intercalation at
the graphene-Ni interface~\cite{marchenko}. A Rashba splitting about
225 meV in epitaxial graphene layers grown on the surface of
Ni~\cite{dedcov} and a giant Rashba SOC ($\sim$ 600 meV) from Pb
intercalation at the graphene-Ir surface~\cite{calleja} are observed
in experiments. In this work, we have studied the interplay between
these two types of SOC on the band structure and the edge
states~\cite{diptiman} with a view to understand the data on charge
conductances.

The electronic properties of GNRs depend on the geometry of the edges
\cite{nakada}, and according to the edge termination type, mainly
there are two kinds of GNR, namely armchair graphene nanoribbon (AGNR)
and zigzag graphene nanoribbon (ZGNR). The ZGNRs are always metallic
with zero band gap, while the AGNRs are metallic when the lateral
width $N = 3M-1$ ($M$ is an integer), else the AGNRs are
semiconducting in nature \cite {fujita} with a finite band gap. The
presence of edges in graphene has strong implications for the
low-energy spectrum of the
$\pi$-electrons~\cite{nakada,fujita,sirgist}. It was shown that ZGNRs
possess localized edge states with energies close to the Fermi
level. On the contrary, edge states are absent for AGNRs. We prefer to
call these GNRs in presence of spin-orbit couplings as Kane-Mele
nanoribbon (KMNR) and we shall deal with the zigzag variant, namely
ZKMNR.

We organize our paper as follows. In the following section, for
completeness and clarity of the system and its notations, we write the
Kane-Mele model for a GNR and perform an analytical investigation of
the edge states for a few choices of the intrinsic and Rashba
spin-orbit coupling. The corresponding band structures are presented
for a check. Subsequently, we include an elaborate discussion of the
results where we have demonstrated the interplay of intrinsic and
Rashba SOC on the charge conductances. We conclude with a brief
summary of our results.

\section{Kane-Mele Hamiltonian}
\label{two}
To begin with a pristine graphene sheet which consists of two
sublattices A and B connected to three nearest neighbors vectors in
real space are given by, $\delta_1 = \big(0, a\big)$; $\delta_2 =
\Big(\frac{\sqrt{3}a}{2}, -\frac{a}{2}\Big)$ and $\delta_3 =
\Big(-\frac{\sqrt{3}a}{2}, -\frac{a}{2}\Big)$, where $a \approx 1.42$
\AA~ is the carbon-carbon distance.  In order to study the nature of the
edge states and transport properties in presence of spin-orbit coupled
term, we consider the following Kane-Mele (KM)
\cite{kane-mele1,kane-mele2} Hamiltonian,
\begin{align}
\label{one} 
H = & - t\sum\limits_{\langle ij\rangle\alpha}c_{i\alpha}^{\dagger}
c_{j\alpha} + it_{2}\sum\limits_{\langle \langle ij
  \rangle\rangle\alpha\beta} \nu_{ij}
c_{i\alpha}^{\dagger}s^z_{\alpha\beta} c_{j\beta} \nonumber \\ & +
i\lambda_R \sum\limits_{\langle ij\rangle\alpha \beta}c_{i\alpha}^{\dagger}
\left( {\bf s} \times {\bf\hat{d}}_{ij}\right)_z c_{j\beta}
\end{align} 
Here $c_{i\alpha}^{\dagger}$ and $c_{j\alpha}$ are creation and
annihilation operators for an electron with spin $\alpha$ on site $i$.
The first term describes the hopping between nearest neighbors $i$ and
$j$ on a honeycomb lattice where the hopping amplitude is, $t$ $\simeq
2.7$ eV. The second term is the mirror symmetric intrinsic spin-orbit
coupling (SOC) which involves spin-dependent second neighbor hopping
between same sublattices with a coupling strength $t_{2}$. $\nu_{ij}$
= $+1(-1)$ if the electron makes a left (right) turn to go from site
$j$ to $i$ through their common nearest neighbor. The vector
\textbf{d} points from site $i$ to site $j$ and corresponds to the
nearest neighbor vectors. $s^z$ is the $z$-component of Pauli spin
matrix. The third term is the nearest neighbor Rashba term which
arises due to the broken surface inversion symmetry with coupling
strength $\lambda_R$. The spin orbit term $t_2$ breaks the $SU(2)$
symmetry down to $U(1)$ symmetry, the Rashba term $\lambda_R$ breaks
the $U(1)$ symmetry down to $\mathbb{Z}_{2}$ \cite{Laubach}. Due to
the small atomic number of carbon, the intrinsic SOC is usually weak
\cite{min, huertas}. The Rashba coupling strength can be tuned by
applying an external electric field.
\begin{figure}[h]
\begin{center}
\includegraphics[width=0.65\textwidth,height=0.45\textwidth]{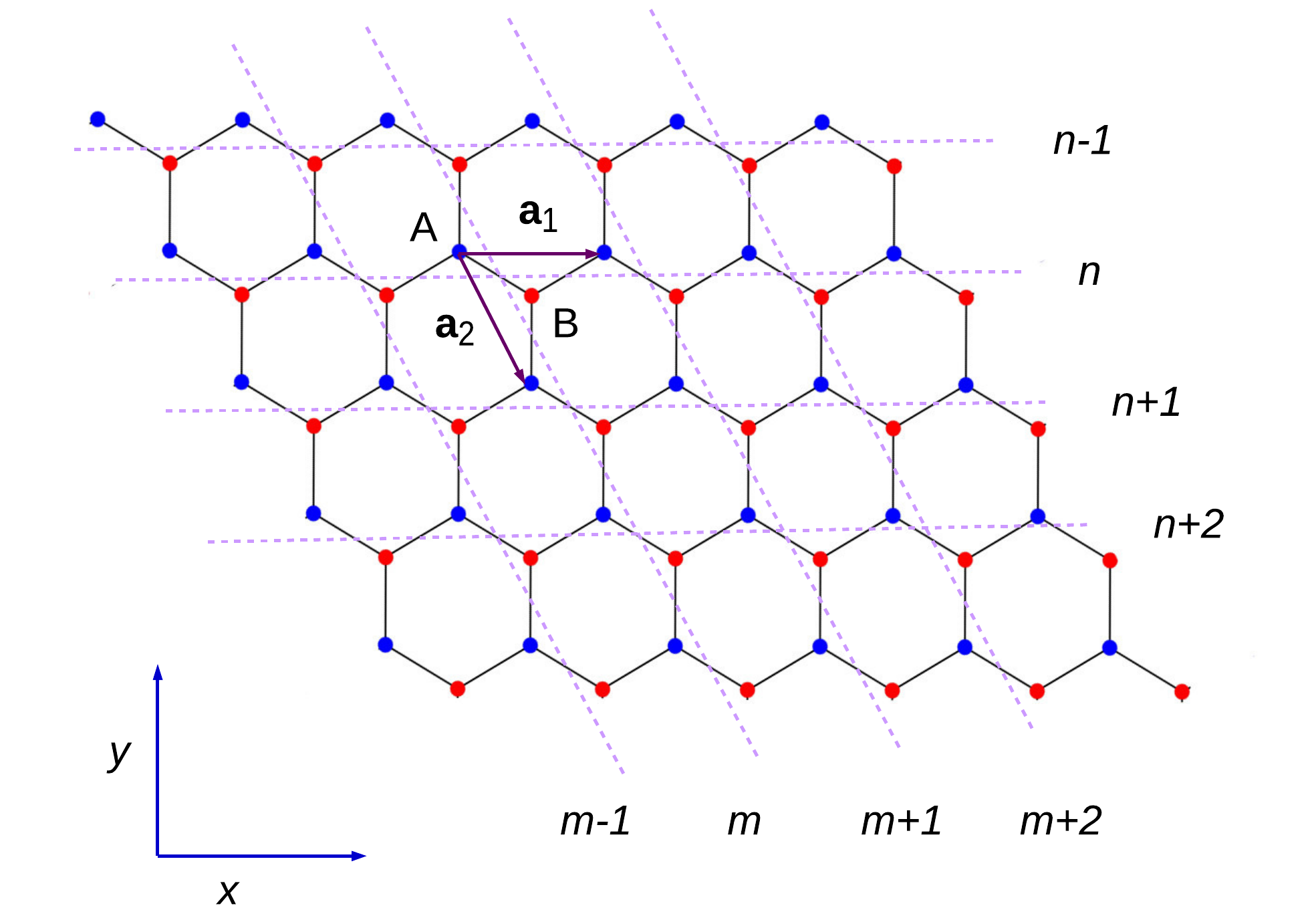}
\caption{(Color online) Graphene nanoribbon geometry with zigzag
  edges. The blue and red circles represent the A and B sublattices of
  the ribbon. $\vec{\bf{a}}_1$ and $\vec{\bf{a}}_2$ are the primitive
  vectors. $(m, n)$ labels the unit cell. $m$ increases along the
  positive $x$-direction whereas $n$ increases along the negative
  $y$-direction.}
\label{ribbon1}
\end{center}
\end{figure}
\section{Edge states}
In this section, we study the edge state properties considering the
tight-binding Hamiltonian in presence of spin-orbit and Rashba
coupling. As said earlier, we focus on Kane-Mele ribbon geometry with
zigzag edges which are infinite along $x$-axis and all the atoms of
the zigzag edges belong to the same sublattice as shown in
Fig.~\ref{ribbon1}. The ribbon width is such that it has $N$ unit
cells in the $y$-axis. We rewrite the equation \eqref{one}, with $t_2$
= $\lambda_R$ = $0$, in terms of $m$, $n$ which labels the unit cell
as shown in Fig.~\ref{ribbon1}, is given by~\cite{neto},
\begin{align} 
H = & - t\sum\limits_{\langle mn
  \rangle\sigma}\Big[a^{\dagger}_{\sigma}(m,n)b_{\sigma}(m,n)+a^{\dagger}_{\sigma}(m,n)b_{\sigma}(m,n-1)
  \nonumber \\ & +a^{\dagger}_{\sigma}(m,n)b_{\sigma}(m-1,n)+ H.C\Big]
\label{h2}
\end{align} 
where $\langle$ $\rangle$ denotes nearest neighbors and $\sigma$
represents the $\uparrow$,$\downarrow$ spin for both sublattices $A$
and $B$.  The translational symmetry exists along $x$-axis but there
is no translational symmetry along the $y$-direction due to the
existence of edge state. Hence the momentum $k_x$ is a good quantum
number. Now we can use a momentum representation of the electron
operator due to the periodicity in the $x$-direction, which is,
\begin{equation}
c_{i} = \frac{1}{\sqrt{L_x}}\sum\limits_{k_x}\exp^{ik_{x} X_{i}} c_{k}
\end{equation}
where ($X_{i}$, $Y_{i}$) represents the coordinate of the site $A$ and $B$ in the unit cell $(m,n)$ (see Fig. \ref{ribbon1}). Using Schr$\ddot{o}$dinger equation $H|\psi\rangle = E |\psi\rangle$, we get the following eigenvalue equations for sites in $A$ and $B$ sublattices as, 
\begin{align}
 &E \alpha\left(k_{x},n\right) = -t\left[2 \cos\left(\frac{\sqrt{3}k_x}{2}\right)\beta\left(k_x,n\right)+\beta\left(k_x, n-1\right)\right] \nonumber
\\
&E \beta\left(k_{x},n\right) = -t\left[2 \cos\left(\frac{\sqrt{3}k_x}{2}\right)\alpha\left(k_x,n\right)+\alpha\left(k_x, n+1\right)\right]
\label{tb_eq}
\end{align}
where we have chosen the basis as, 
\begin{equation}
|\psi_{k}\rangle=\sum\limits_{n, \sigma}\alpha(k, n, \sigma)|a, k, n, \sigma\rangle+\beta(k, n, \sigma)|b, k, n, \sigma\rangle. \nonumber
\end{equation}
where $\alpha$, $\beta$ are the eigenstates for the $A$ and $B$ sublattices. Since we have taken zigzag edges and the ribbon exists only between $0$ to $N-1$, and thus the boundary condition is,
\begin{equation}
\alpha(k_x,n)= \beta(k_x,-1)=0
\label{bc}
\end{equation}
and we impose $E =0$ as the plateau in the conductance is observed for
$E \simeq 0$. We have made $k_x$ as a dimensionless quantity by
absorbing the lattice spacing $a$ into the definition of $k_x$. We can
see that the charge density is proportional to
$\left[2\cos\left(\frac{\sqrt{3}k_x}{2}\right)\right]^{n}$ at each
non-nodal site of the $n$-th zigzag chain certifying the presence of
localized edge states with an exponential decay as one deviates from
the edge\cite{nakada}. A completely flat band exists in the special
vicinity of $k_x$. Since the cosine term becomes zero at $k_x =
\frac{\pi}{\sqrt{3}}$, we have taken $k_x = \frac{n\pi}{3\sqrt{3}}$ \cite{Priyanka} to
arrive the desired solution for a pristine graphene.

Now we consider the intrinsic spin-orbit interaction term as mentioned
in Eq.~\eqref{one}. Using
the same basis and including spin degrees of freedom we get four
eigenvalue equations for spins up and down and for $A$ and $B$
sublattice points which are written as,
%
\begin{align}
E\alpha_{\uparrow}\left(k_{x},n\right) =&
-t\left[2\cos\left(\frac{\sqrt{3}k_x}{2}\right)\beta_{\uparrow}(k_x,n)
  + \beta_{\uparrow}(k_x, n-1)\right] \nonumber \\ & - 2
t_2\Bigg[\sin\left(\sqrt{3}k_x\right)\alpha_{\uparrow}(k_x,n)-
  \sin\left(\frac{\sqrt{3}k_x}{2}\right)\bigg(\alpha_{\uparrow}(k_x,n-1)+\alpha_{\uparrow}(k_x,n+1)\bigg)\Bigg],
\nonumber \\ E\alpha_{\downarrow}\left(k_{x},n\right) = & -t\left[2
  \cos\left(\frac{\sqrt{3}k_x}{2}\right)\beta_{\downarrow}(k_x,n) +
  \beta_{\downarrow}(k_x, n-1)\right] \nonumber \\ & + 2
t_2\Bigg[\sin\left(\sqrt{3}k_x\right)\alpha_{\downarrow}(k_x,n)-\sin\left(\frac{\sqrt{3}k_x}{2}\right)\bigg(\alpha_{\downarrow}(k_x,n-1)
  + \alpha_{\downarrow}(k_x,n+1)\bigg)\Bigg], \nonumber
\\ E\beta_{\uparrow}\left(k_{x},n\right) =& -t\left[2
  \cos\left(\frac{\sqrt{3}k_x}{2}\right)\alpha_{\uparrow}(k_x,n) +
  \alpha_{\uparrow}(k_x, n+1)\right] \nonumber \\ & +
2t_2\Bigg[\sin\left(\sqrt{3}k_x\right)\beta_{\uparrow}(k_x,n)-\sin\left(\frac{\sqrt{3}k_x}{2}\right)\bigg(\beta_{\uparrow}(k_x,n-1)+\beta_{\uparrow}(k_x,n+1)\bigg)\Bigg],\nonumber
\\ E\beta_{\downarrow}\left(k_{x},n\right) =& - t\left[2
  \cos\left(\frac{\sqrt{3}k_x}{2}\right)\alpha_{\downarrow}(k_x,n) +
  \alpha_{\downarrow}(k_x, n+1)\right] \nonumber \\ & - 2
t_2\Bigg[\sin\left(\sqrt{3}k_x\right)\beta_{\downarrow}(k_x,n)-\sin\left(\frac{\sqrt{3}k_x}{2}\right)\bigg(\beta_{\downarrow}(k_x,n-1)+\beta_{\downarrow}(k_x,n+1)\bigg)\Bigg],
\label{tripple_eqn1}
\end{align}
where $\alpha_{\uparrow,\downarrow}$ and $\beta_{\uparrow,\downarrow}$
correspond to the spin resolved eigenstates for the $A$ and $B$
sublattices.  For a comprehensible solution, we turn to a numerical
computation of the above set of equations. Here we have used $k_x =
\frac{\pi}{\sqrt{3}}$ which particularly renders simple forms for the
equations. Solving the above equations (Eq.~\ref{tripple_eqn1})
numerically (with $t =1$ and energy $E =0$) following the same
boundary conditions (as given in Eq.~\ref{bc}), we shall see how the
probability densities of the wavefunctions decay ascertaining whether
edge states exist.

Next we turn on the other SOC, namely the Rashba spin-orbit coupling
(RSOC). This yields the new set of equations for a KMNR given by
\cite{diptiman2},
\begin{eqnarray} 
E\alpha_{\uparrow}\left(k_{x},n\right) =&
-&t\left[2\cos\left(\frac{\sqrt{3}k_x}{2}\right)\beta_{\uparrow}(k_x,n)
  + \beta_{\uparrow}(k_x, n-1)\right]\nonumber \\ & -&2
t_2\Bigg[\sin\left(\sqrt{3}k_x\right)\alpha_{\uparrow}(k_x,n)-\sin\left(\frac{\sqrt{3}k_x}{2}\right)\bigg(\alpha_{\uparrow}(k_x,n-1)+\alpha_{\uparrow}(k_x,n+1)\bigg)\Bigg]\nonumber
\\ &
+&i\lambda_R\left[\left(\cos\left(\frac{\sqrt{3}k_x}{2}\right)+\sqrt{3}\sin\left(\frac{\sqrt{3}k_x}{2}\right)\right)\beta_\downarrow(k_x,
  n)-\beta_\downarrow(k_x,n-1)\right], \nonumber
\\
 E\alpha_{\downarrow}\left(k_{x},n\right) = & -&t\left[2
  \cos\left(\frac{\sqrt{3}k_x}{2}\right)\beta_{\downarrow}(k_x,n) +
  \beta_{\downarrow}(k_x, n-1)\right] \nonumber \\ & +& 2
t_2\Bigg[\sin\left(\sqrt{3}k_x\right)\alpha_{\downarrow}(k_x,n)-\sin\left(\frac{\sqrt{3}k_x}{2}\right)\bigg(\alpha_{\downarrow}(k_x,n-1)+\alpha_{\downarrow}(k_x,n+1)\bigg)\Bigg]\nonumber
\\ &
+&i\lambda_R\left[\left(\cos\left(\frac{\sqrt{3}k_x}{2}\right)-\sqrt{3}\sin\left(\frac{\sqrt{3}k_x}{2}\right)\right)\beta_\uparrow(k_x,
  n)-\beta_\uparrow(k_x,n-1)\right], \nonumber
\end{eqnarray} 
\begin{eqnarray}
 E\beta_{\uparrow}\left(k_{x},n\right) =& -&t\left[2
  \cos\left(\frac{\sqrt{3}k_x}{2}\right)\alpha_{\uparrow}(k_x,n) +
  \alpha_{\uparrow}(k_x, n+1)\right]\nonumber \\ & +& 2
t_2\Bigg[\sin\left(\sqrt{3}k_x\right)\beta_{\uparrow}(k_x,n)-\sin\left(\frac{\sqrt{3}k_x}{2}\right)\bigg(\beta_{\uparrow}(k_x,n-1)+\beta_{\uparrow}(k_x,n+1)\bigg)\Bigg]
\nonumber \\ &
-&i\lambda_R\left[\left(\cos\left(\frac{\sqrt{3}k_x}{2}\right)-\sqrt{3}\sin\left(\frac{\sqrt{3}k_x}{2}\right)\right)\alpha_\downarrow(k_x,
  n)-\alpha_\downarrow(k_x,n+1)\right], \nonumber
\\ E\beta_{\downarrow}\left(k_{x},n\right) =& -& t\left[2
  \cos\left(\frac{\sqrt{3}k_x}{2}\right)\alpha_{\downarrow}(k_x,n) +
  \alpha_{\downarrow}(k_x, n+1)\right] \nonumber \\ & -& 2
t_2\Bigg[\sin\left(\sqrt{3}k_x\right)\beta_{\downarrow}(k_x,n)-\sin\left(\frac{\sqrt{3}k_x}{2}\right)\bigg(\beta_{\downarrow}(k_x,n-1)+\beta_{\downarrow}(k_x,n+1)\bigg)\Bigg]
\nonumber \\ &
-&i\lambda_R\left[\left(\cos\left(\frac{\sqrt{3}k_x}{2}\right)+\sqrt{3}\sin\left(\frac{\sqrt{3}k_x}{2}\right)\right)\alpha_\uparrow(k_x,
  n)-\alpha_\uparrow(k_x,n+1)\right].
\label{tripple_eqn2}
\end{eqnarray}
It can be checked that for $\lambda_R = 0$ we retrieve
Eq. \ref{tripple_eqn1}. It is clear from the above equations that
$s_z$ is non-conserved and the spin of the edges can be rotated
\cite{strained}. We define the total probability as,
\begin{equation}
|\psi|^2 =\sum\limits_{\sigma} |\psi_{A\sigma}|^2 + |\psi_{B\sigma}|^2
\end{equation}
The probability of finding an electron in the spin-up state that is,
$|\psi_{A\uparrow}|^{2} + |\psi_{B\uparrow}|^{2}$ is equal to the
probability of finding an electron in the spin-down state that is,
$|\psi_{A\downarrow}|^{2} + |\psi_{B\downarrow}|^{2}$. This is the
evidence that the RSOC does not break time-reversal symmetry
\cite{sandler}.

\section{Transport properties of KMNR}
The electronic transport properties of KMNR show interesting phenomena
due to the presence of their edge states. To study the electron
conductance we investigate the transport characteristics using
Landauer-B\"uttiker formula \cite {land_cond,land_cond2} that relates
the scattering matrix to the conductance via,
\begin{equation}
G = \frac{e^2}{h} T(E)
\end{equation}
where $T(E)$ is the transmission coefficient.The transmission
coefficient is defined as \cite{caroli,Fisher-Lee},
\begin{equation}
T = \text{Tr}\left[\Gamma_R {\cal G}_R
  \Gamma_L {\cal G}_A\right]
\end{equation}
${\cal G}_{R(A)}$ is the retarded (advanced) Green's function of the
scattering region. The coupling matrices $\Gamma_{L(R)}$ are the
imaginary parts representing the coupling between the scattering
region and the left (right) lead. They are defined by \cite{dutta},
\begin{equation}
\Gamma_{L(R)} = i\left[\Sigma_{L(R)} -
  (\Sigma_{L(R)})^\dagger\right]
\end{equation}
Here $\Sigma_{L(R)}$ is the retarded self-energy associated with the
left (right) lead. The self-energy contribution is computed by
modeling each terminal as a semi-infinite perfect wire
\cite{nico}. From Green's function , the local density of states
(LDOS) can be found \cite{dutta},
\begin{equation}
\rho(E) = -\frac{1}{\pi} Im \left[G(E)\right]
\end{equation}
To compute the LDOS maps, the retarded Greens function can be used for
the system is written as\cite{dutta},

\begin{equation}
G(E) = \frac{1}{EI-H+i\eta}
\end{equation}

\section{Results and discussion}
To begin with, let us discuss the values of the spin-orbit couplings
used in our work. Ideally, the strengths of both kinds of SOC are too
weak to observe perceptible effects. For example gold (Au) and
Thallium (Tl) decorated GNRs yield the following values for the SOC,
namely $t_2 = 0.007$, $\lambda_R = 0.0165$ and $t_2 = 0.02$,
$\lambda_R =0$ respectively (all quoted in units of hopping,
$t$). However in our work, without much trepidation, we take
$\lambda_R$ and $t_2$ as parameters (this also provides a
justification in calling the system as Kane-Mele nanoribbon
(KMNR)). Here we have taken $t_2 = 0.1$ and 0.5 and considered
different values of $\lambda_R$ in the range [0.01 : 0.5]. We have
also examined other values of $t_2$ and $\lambda_R$, however, they do
not produce any qualitatively new results than the ones already
presented in this work.

Our focus is to understand the nature of the edge states via both
analytic and numeric computations and their effects on the conductance
spectra of a KMNR. To distinguish between various cases, we have
considered (a) a pristine GNR, with $t_2 = \lambda_R= 0$, (b) KMNR
with only intrinsic SOC, that is, $t_2 \neq 0$ and $\lambda_R= 0$ (as
is the case for Tl decorated GNR, albeit with an overestimated $t_2$),
(c) KMNR with only Rashba SOC, $t_2= 0$, $\lambda_R \neq 0$, and (d)
KMNR with $t_2 \neq 0$, $\lambda_R \neq 0$. Further, to have a lucid
visualization of the existence of edge states and compare with the
results obtained above we plot the band structure in each of the
cases.

A bit of details on our numeric computation can be given as
follows. We have taken $N = 100$ unit cells in the $y$-direction (see
Fig. \ref{ribbon1}) and thus the Hamiltonian in Eq. \ref{one} is a
$400 \times 400$ matrix owing to both spin and sublattice degrees of
freedom. We have set the tight-binding parameter, $t=1$ and the
lattice spacing, $a=1$. All the energies are measured in units of
$t$. For our numerical calculation on the LDOS and conductance we have
used KWANT \cite{kwant}. The size of the KMNR in numeric computation
is taken as 81Z-40A~\cite{sudin-mrx} with zigzag edges.
\subsection{Pristine graphene}
\label{pristine}
\begin{figure}[!ht]
 \centering \subfloat{\includegraphics[trim=0 0 0
     0,clip,width=0.5\textwidth]{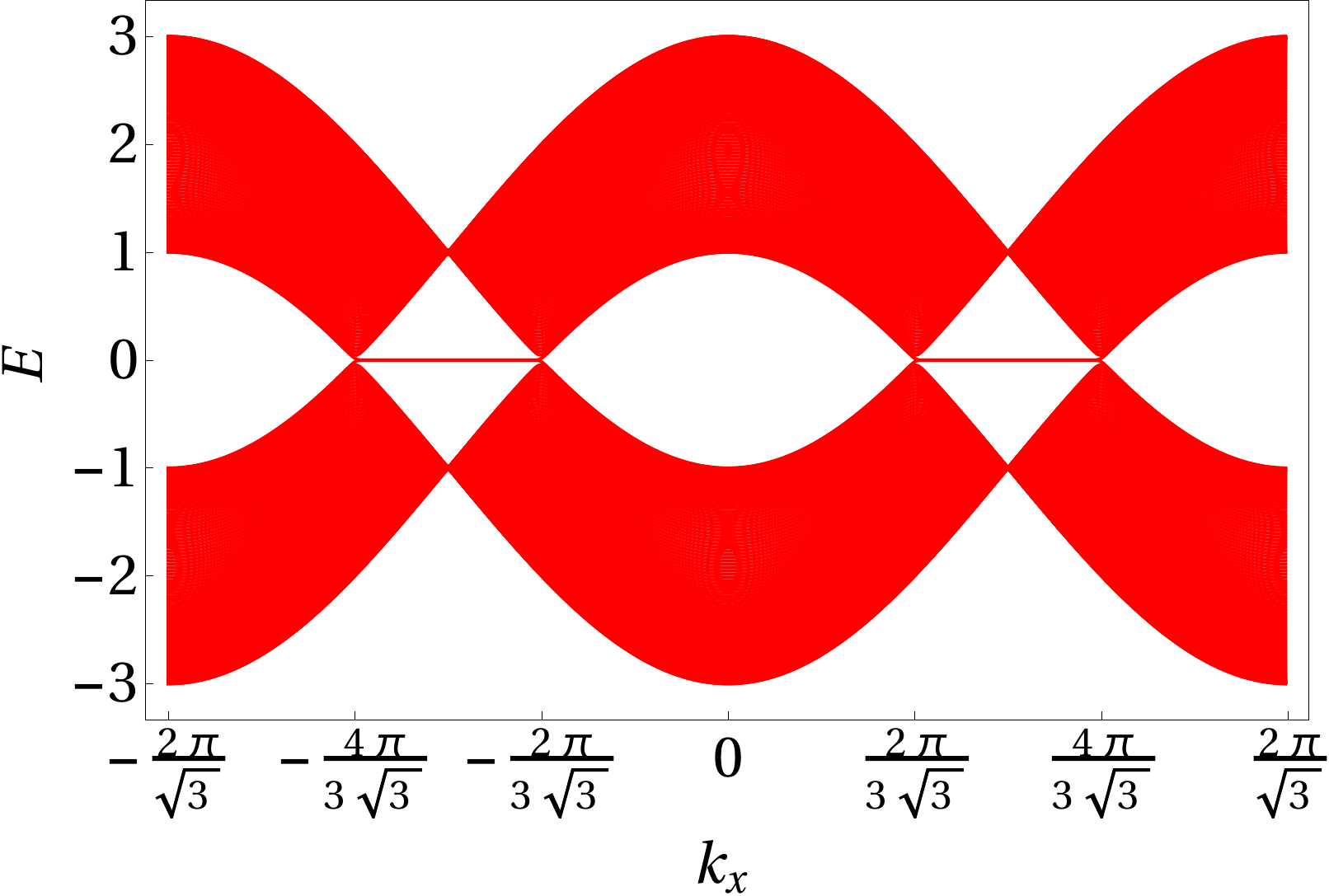}\label{fig:2}}
  \caption{(color online) Band structure of zigzag ribbon for a pristine graphene.}
\label{pristine_gr1}
\end{figure}

In pristine graphene, we put $t_2 = \lambda_R =0$ in
Eq. \ref{one}. The computed band structure is presented in
Fig.~\ref{pristine_gr1}. It can be easily observed that the flat band
exists at exactly zero energy lying between the values of $k_x = -4
\pi/3\sqrt{3}$ and $-2\pi /3\sqrt{3}$ and between $k_x = 4
\pi/3\sqrt{3}$ and $2\pi /3\sqrt{3}$ \cite {diptiman}. These bands
represent that the states are localized at the edges \cite{lado}.

\begin{figure}[!ht]
 \centering
   \subfloat{\includegraphics[trim=0 0 0 0,clip,width=0.4\textwidth,height=0.35\textwidth]{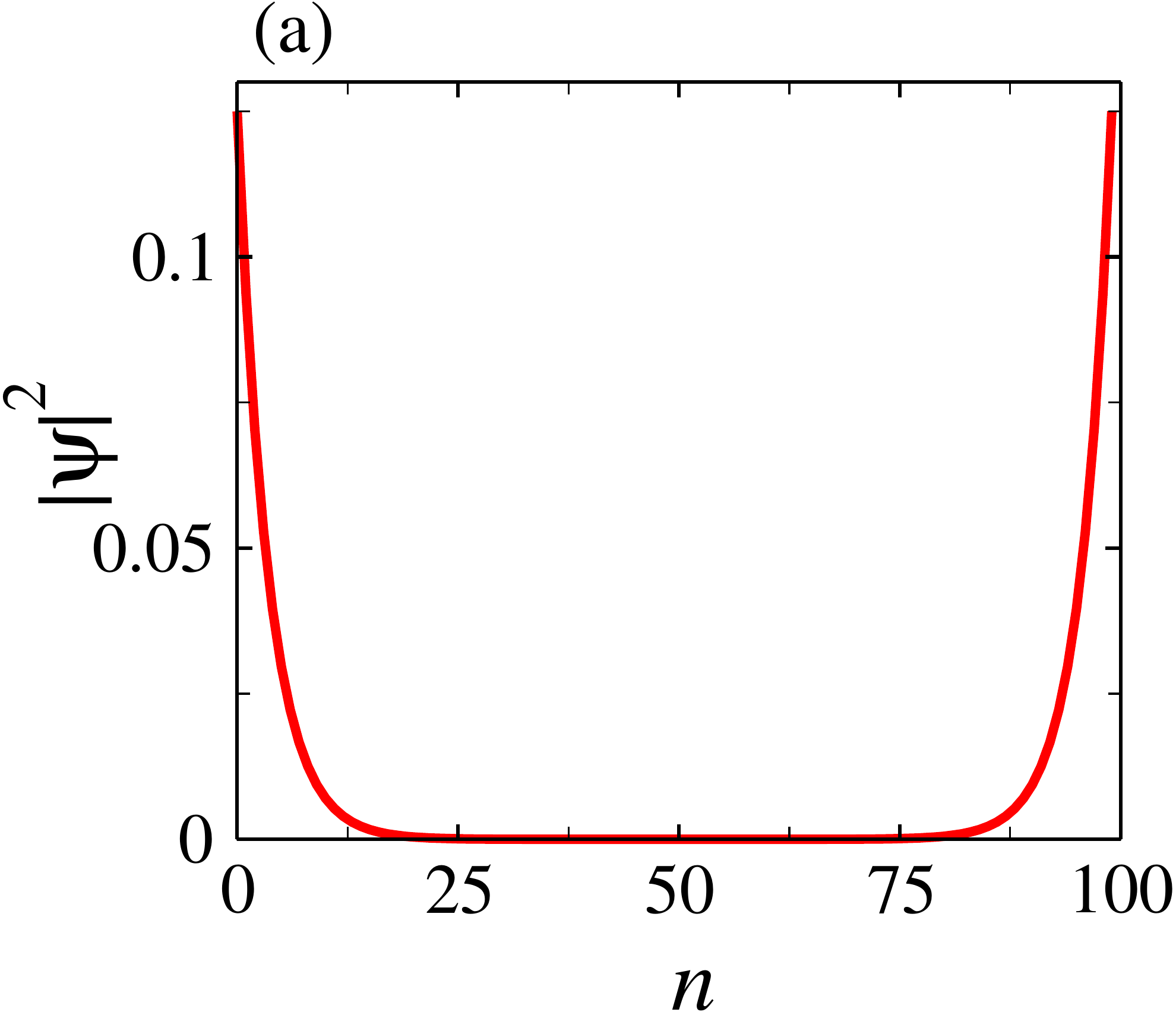}\label{fig:3a}} \hspace{0.2 cm}
   \subfloat{\includegraphics[trim=0 -23 0 0,clip, width=0.4\textwidth,height=0.35\textwidth]{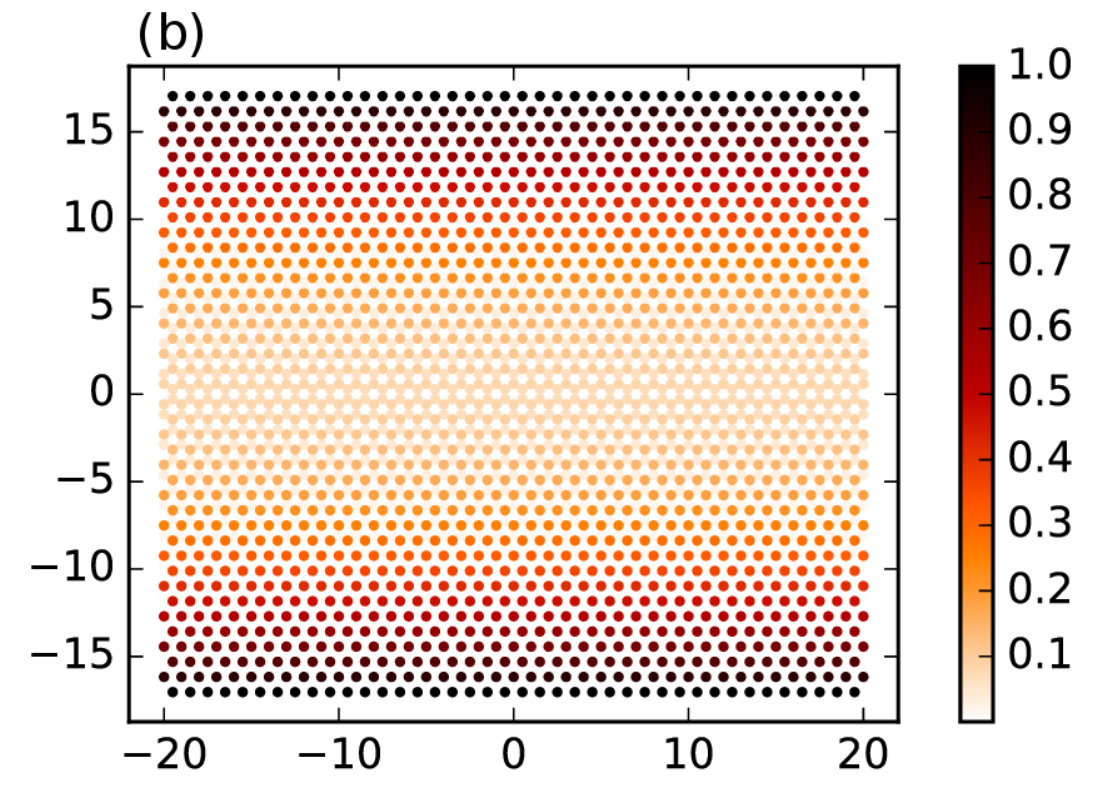}\label{fig:3b}}  
  \caption{(color online) (a) Probability density of the wavefunction,
    $|\psi|^{2}$ as a function of $n$ and (b) the LDOS
    plot for pristine graphene.}
\label{pristine_gr2}
\end{figure}

To see how the states at edges look like, we have plotted the
probability density, $|\psi|^2$ as a function $n$
by solving the Eq.~\ref{tb_eq} and present them in
Fig.~\ref{pristine_gr2} (a). It shows that the wavefunction has
maximum amplitudes at the edges and gradually decays as one moves
inwards. The lack of a gap in pristine graphene turns the $E= 0 $
state into a resonance whose amplitude decays \cite{lado}. To compare
 the analytical results with numerical ones, we have plotted
the local density of states (LDOS) in Fig.~\ref{pristine_gr2} (b) for
energy value close to zero. It can be seen that the LDOS is largest at
the edges and falls off gradually into the bulk. Thus these edge
states are conducting in nature, whereas deep inside the bulk remains
insulating owing to its decaying amplitude.



\begin{figure}[!ht]
 \centering
   \subfloat{\includegraphics[trim=0 0 0 0,clip,width=0.5\textwidth]{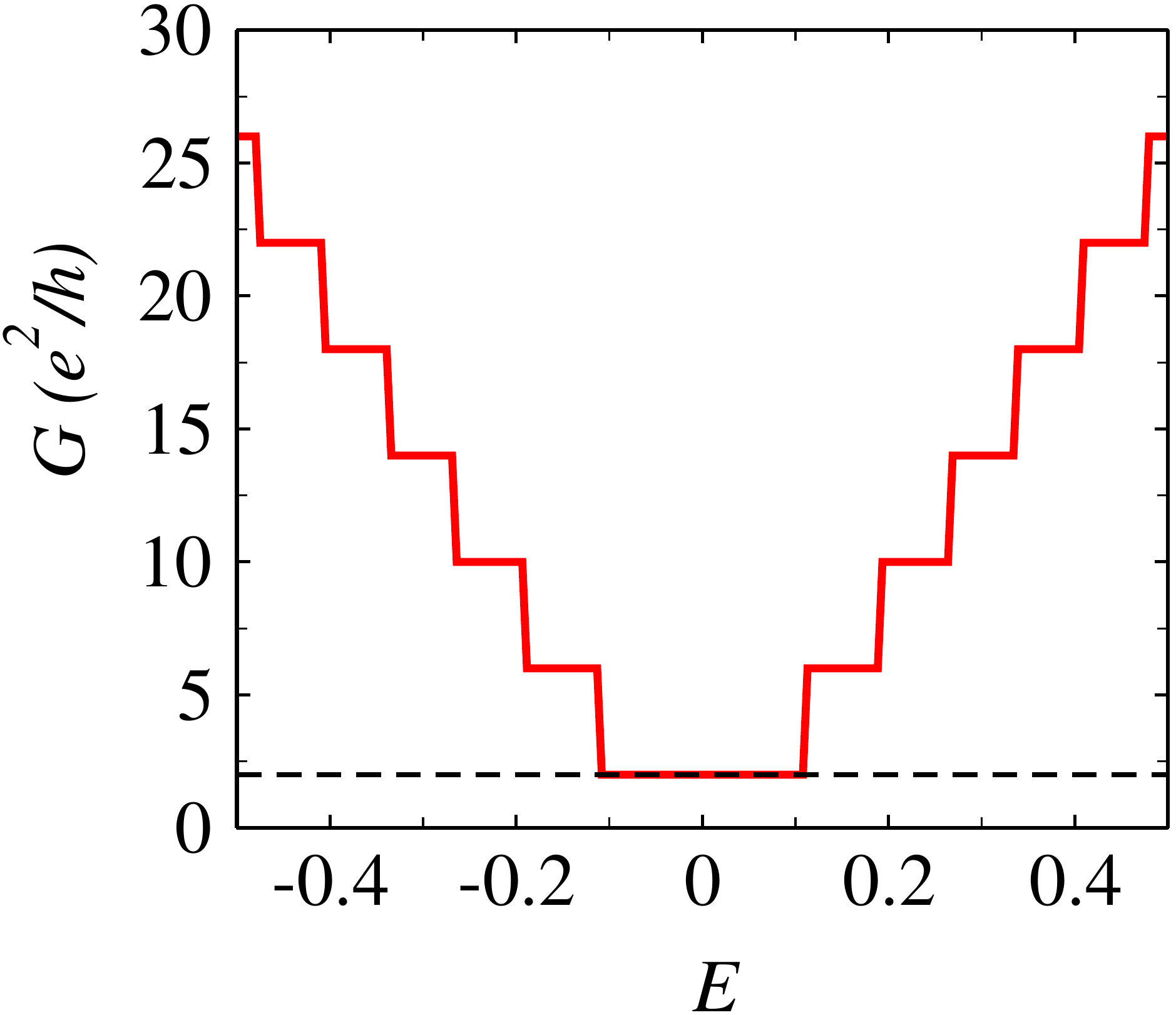}\label{fig:4}} 
  \caption{(color online) The charge conductance, $G$ (in units of
    $e^2/h$) is plotted as a function of energy $E$ (in units of $t$)
    for pristine graphene.}
\label{pristine_gr3}
\end{figure}

To ascertain the effects of edge modes on the conductance properties
of a KMNR, we have computed the conductance of pristine (zigzag)
graphene as shown in Fig. \ref{pristine_gr3}. The conductance behavior
shows that a $2e^2/h$ plateau exists around the zero of the Fermi
energy which is shown by the black dotted line. However, this plateau is
fragile owing to the absence of `protected' edge states.

\subsection{Intrinsic SOC}
\label{SOC}

\begin{figure}[!ht]
  \centering
  \subfloat[]{\includegraphics[trim=0 0 0 0,clip,width=0.45\textwidth]{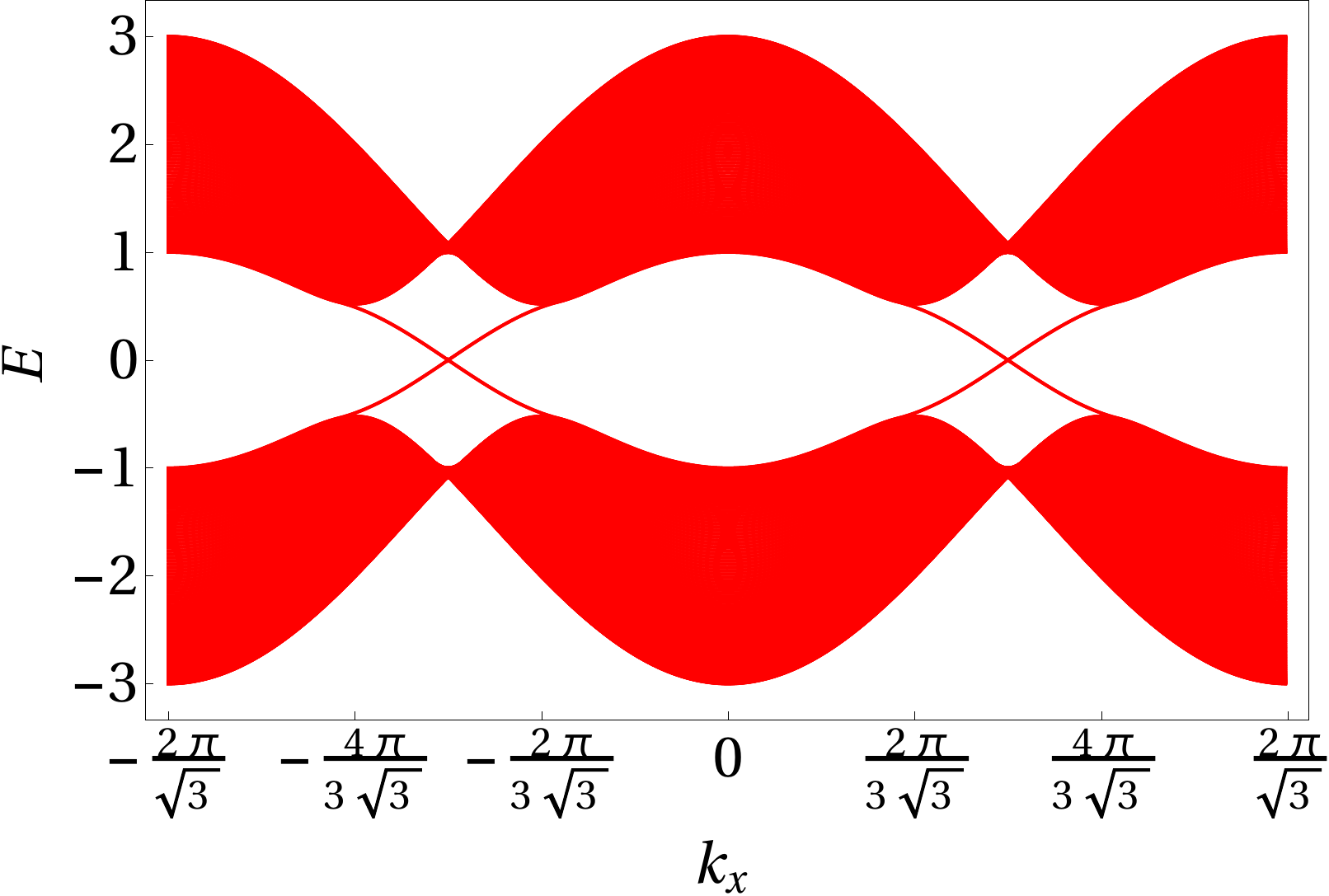}\label{fig:5a}} \hspace{0.2 cm}
  \subfloat[]{\includegraphics[trim=0 0 0 0,clip,width=0.45\textwidth]{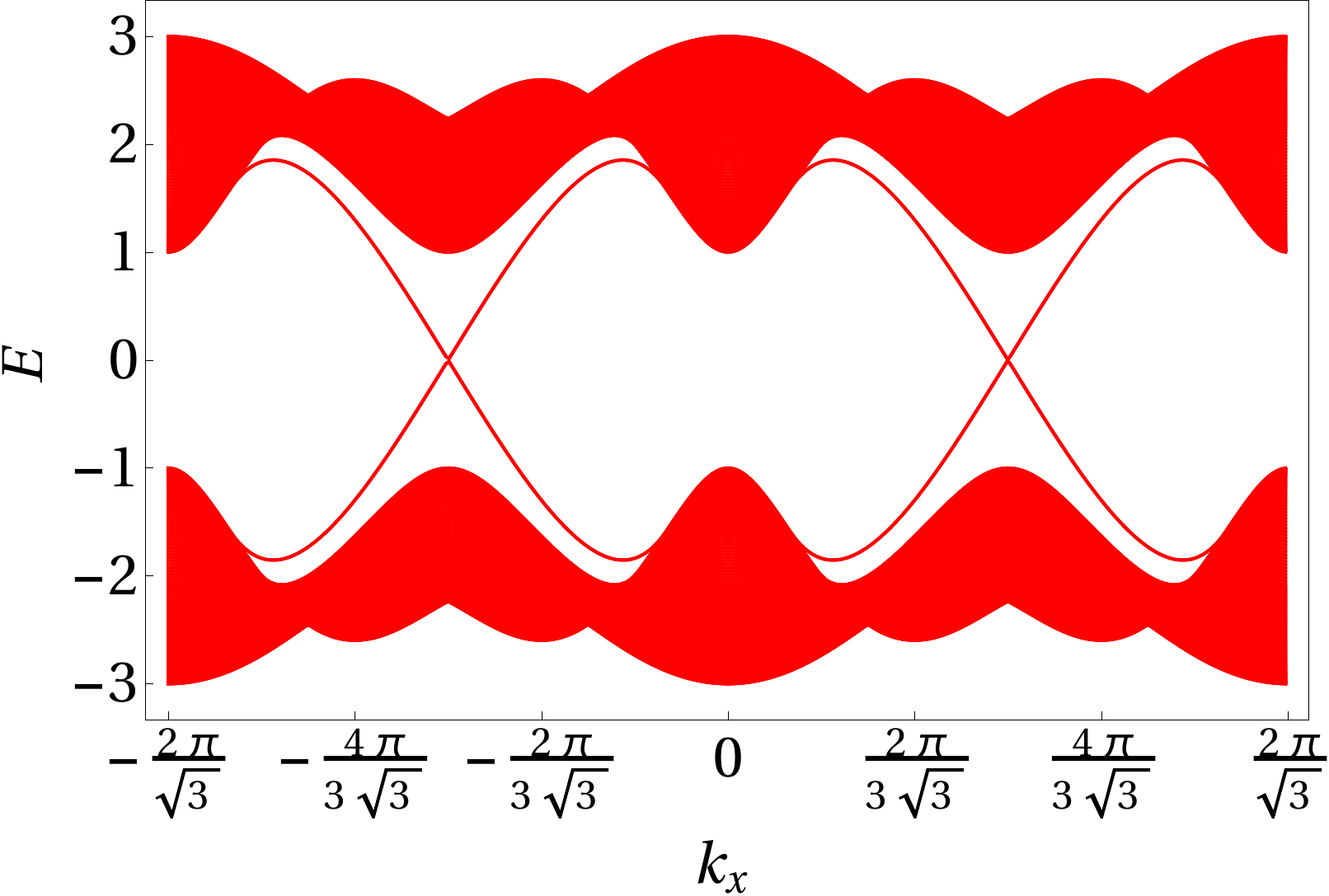}\label{fig:5b}} 
  \caption{(color online) Band structure of zigzag graphene ribbon with coupling parameter (a) $t_2 = 0.1$ and (b) $t_2 = 0.5$.}
\label{soc1}
\end{figure}

\begin{figure}[!ht]
  \centering \subfloat{\includegraphics[trim=0 0 0
      0,clip,width=0.4\textwidth,height=0.35\textwidth]{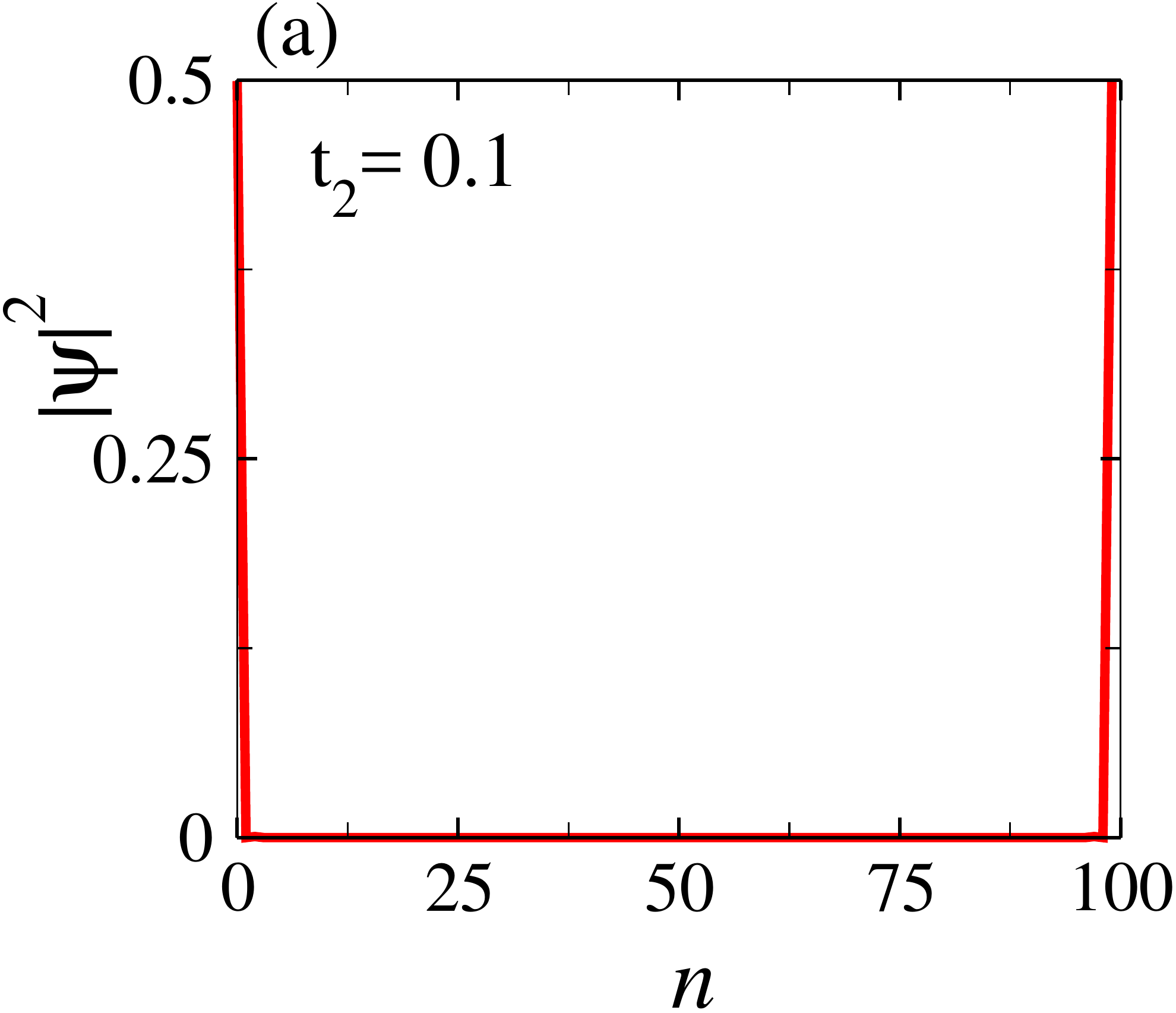}\label{fig:6a}}\hspace{0.2
    cm} \subfloat{\includegraphics[trim=0 -23 0
      0,clip,width=0.4\textwidth,height=0.35\textwidth]{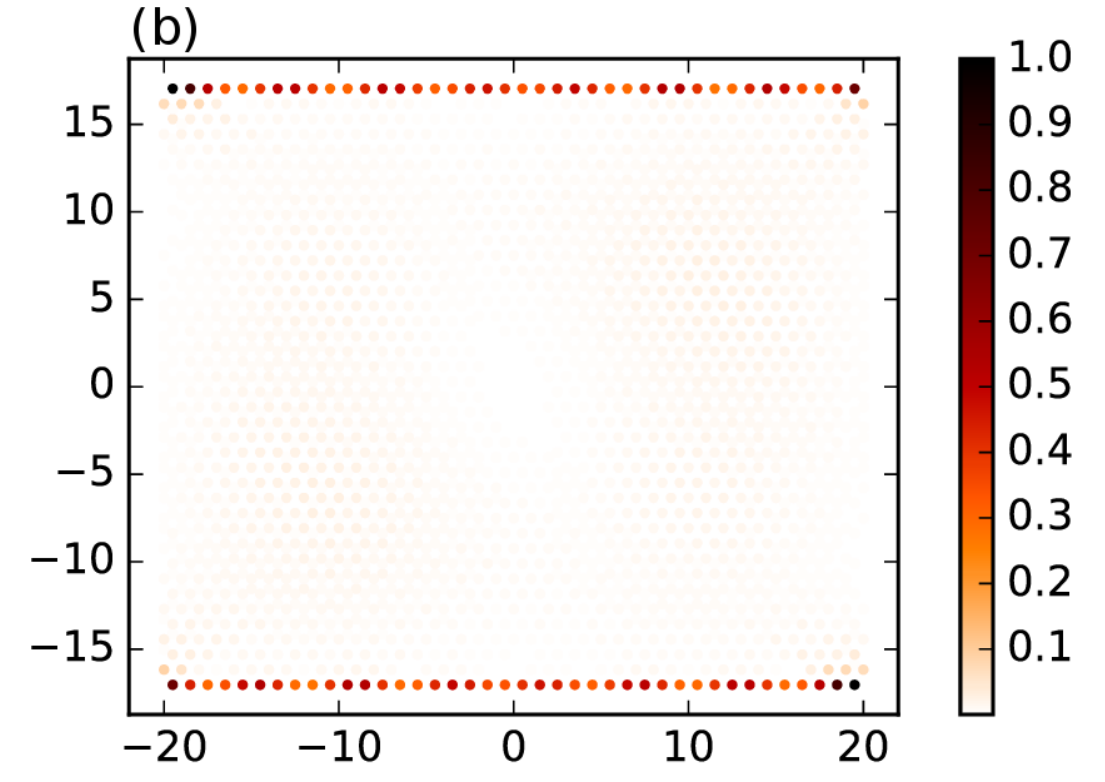}\label{fig:6b}}
  \\ \subfloat{\includegraphics[trim=0 0 0
      0,clip,width=0.4\textwidth,height=0.35\textwidth]{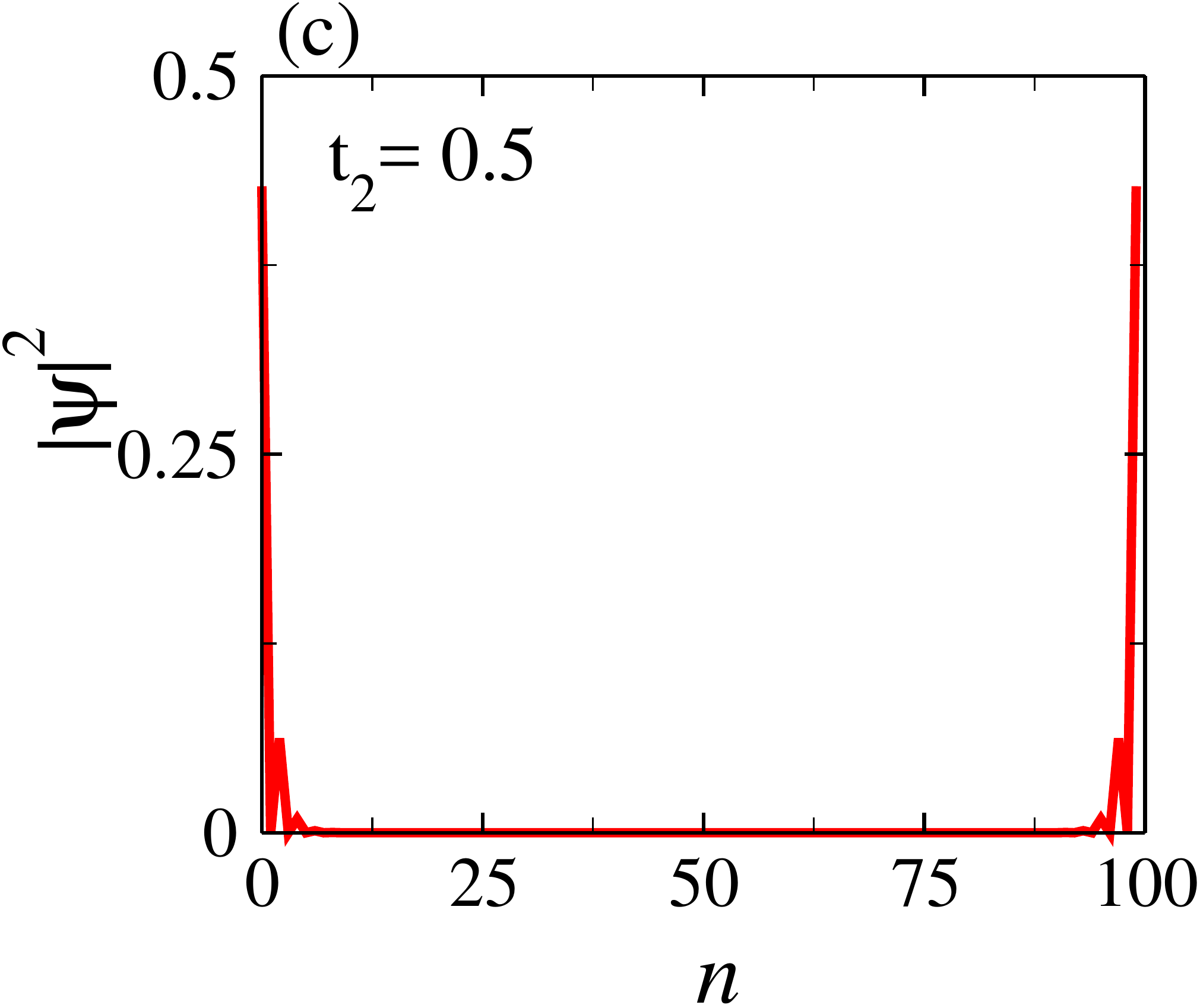}\label{fig:6c}}\hspace{0.2
    cm} \subfloat{\includegraphics[trim=0 -24 0
      0,clip,width=0.4\textwidth,height=0.35\textwidth]{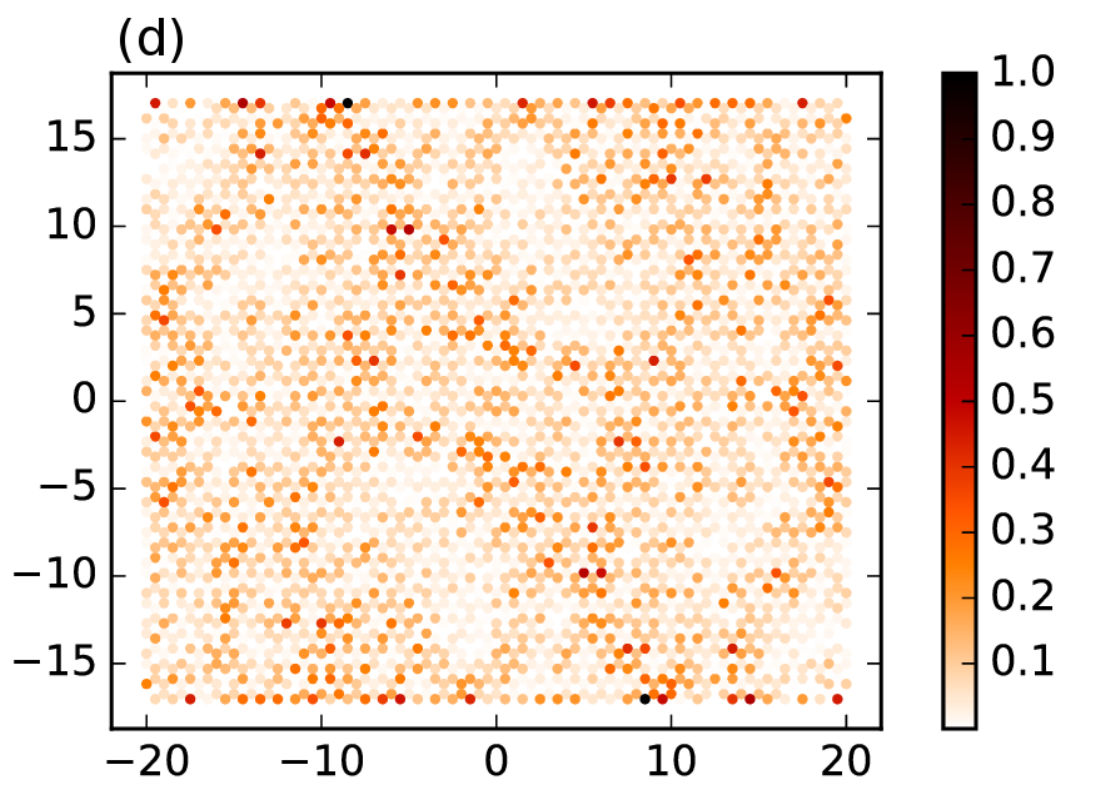}\label{fig:6d}}
  \caption{(Color online) Probability density of the wave-function,
    $|\psi|^{2}$ as a function of $n$ for (a) $t_2=0.1$
    and (c) $t_2=0.5$. The LDOS plot for (b) $t_2=0.1$ and (d)
    $t_2=0.5$. Here $\lambda_R$ is not considered.}
\label{soc2}
\end{figure}

In this subsection, we shall discuss results for an intrinsic SOC
added to a pristine graphene. Fig.~\ref{soc1} shows the band structure
with two different intrinsic SO coupling strengths, namely $t_2 = 0.1$
and 0.5. The system dimensions have been kept the same as that in pristine
graphene. Now from fig. \ref{soc1} (a), we see a gap has opened at the
Dirac points which may appear that the system is a spin Hall insulator
(SHI) \cite{murakami} with edge modes. As shown in Fig.~\ref{soc1}
(b), a larger gap than the previous case is observed. To
confirm the existence of edge modes, we have also plotted the
probability density and local density of states (LDOS) as shown in
Fig.~\ref{soc2}.


We have plotted the probability density for the edge states with the
strength of the intrinsic coupling, $t_2 = 0.1$ as shown in
Fig.~\ref{soc2} (a). Now the edge states fall off sharply on both
sides of the sample. We have also plotted the LDOS for the
comparison. The corresponding Fig.~\ref{soc2} (b) implies that the
electronic states are highly localized at the edges and are zero
immediately inwards. However, the inclusion of the intrinsic SOC respects the time reversal symmetry of the Kane-Mele Hamiltonian and
hence these edge states should be protected by
topological invariance. We have also plotted the probability
amplitude in Fig.~\ref{soc2} (c) and the LDOS in Fig.~\ref{soc2} (d)
for a larger intrinsic SOC, namely $t_2=0.5$ . The probability
amplitude now does not decay sharply as that for $t_2=0.1$ and also
this result is in agreement with the LDOS plot where the states are
not localized at the edges of the sample rather show an oscillating
nature. The
localized edge states as seen from Fig.~\ref{soc2} (a) and
Fig.~\ref{soc2} (b) conduct and should yield a non-zero conductance
value at zero bias.

\begin{figure}[!ht]
  \centering \subfloat{\includegraphics[trim=0 0 0
      0,clip,width=0.4\textwidth,height=0.35\textwidth]{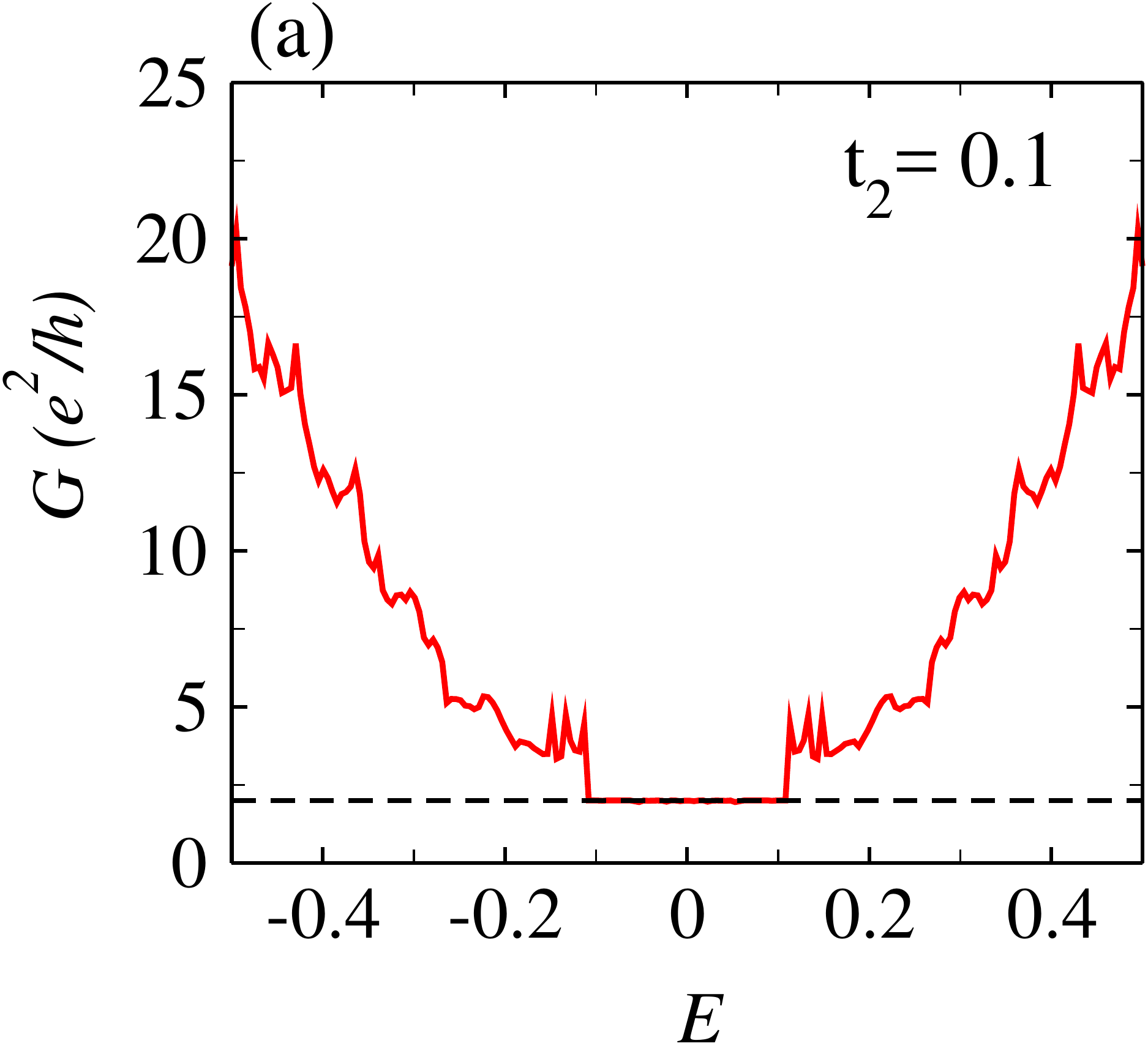}\label{fig:7a}} \hspace{0.2
    cm} \subfloat{\includegraphics[trim=0 0 0
      0,clip,width=0.4\textwidth,height=0.35\textwidth]{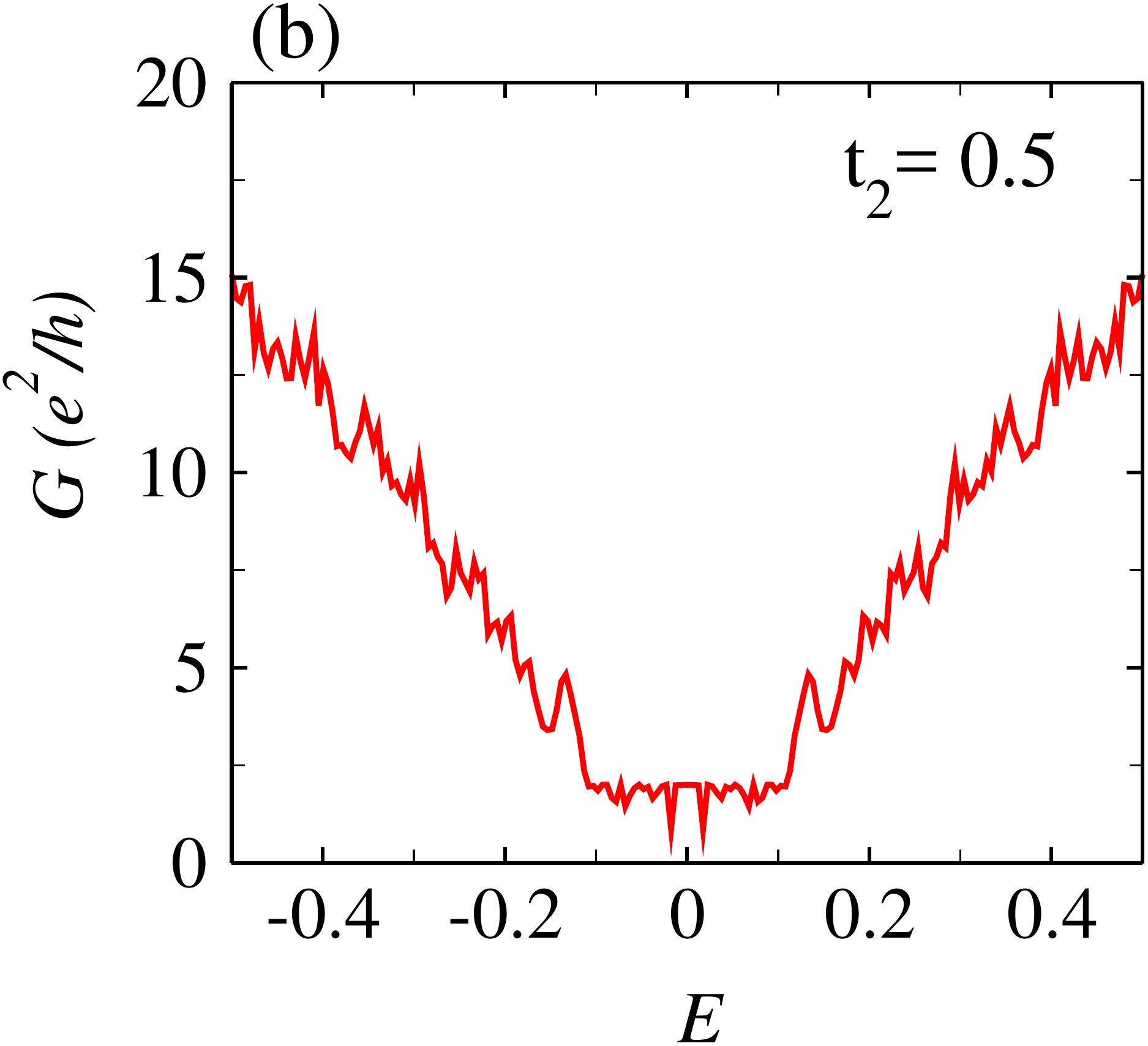}\label{fig:7b}}
  \caption{(color online) The charge conductance, $G$ (in units of
    $e^2/h$) is plotted as a function of energy E (in units of $t$)
    for (a) $t_2 = 0.1$ and (b) $t_2 = 0.5$. We have considered
    $\lambda_R=0$}
\label{soc3}
\end{figure}

To see this we have plotted the (charge) conductance, $G$ as a
function of Fermi energy, $E$ in presence of the intrinsic SOC as
shown in Fig.~\ref{soc3}. Although the step-like behavior is absent
unlike that of pristine graphene, a $2e^2/h$ plateau is still observed
for the case of $t_2=0.1$ shown by the black dotted line in
Fig.~\ref{soc3} (a). However, for $t_2=0.5$ as shown in
Fig.~\ref{soc3} (b), there is no $2e^2/h$ plateau near the zero of the
Fermi energy. It is also important to note that the maximum value of
the conductance, that is at $|E|\simeq0.5$ is higher for lower values
of the intrinsic SOC.


\subsection{Rashba SOC}
\label{RSOC}
\begin{figure}[!ht]
  \centering \subfloat[]{\includegraphics[trim=0 0 0
      0,clip,width=0.45\textwidth]{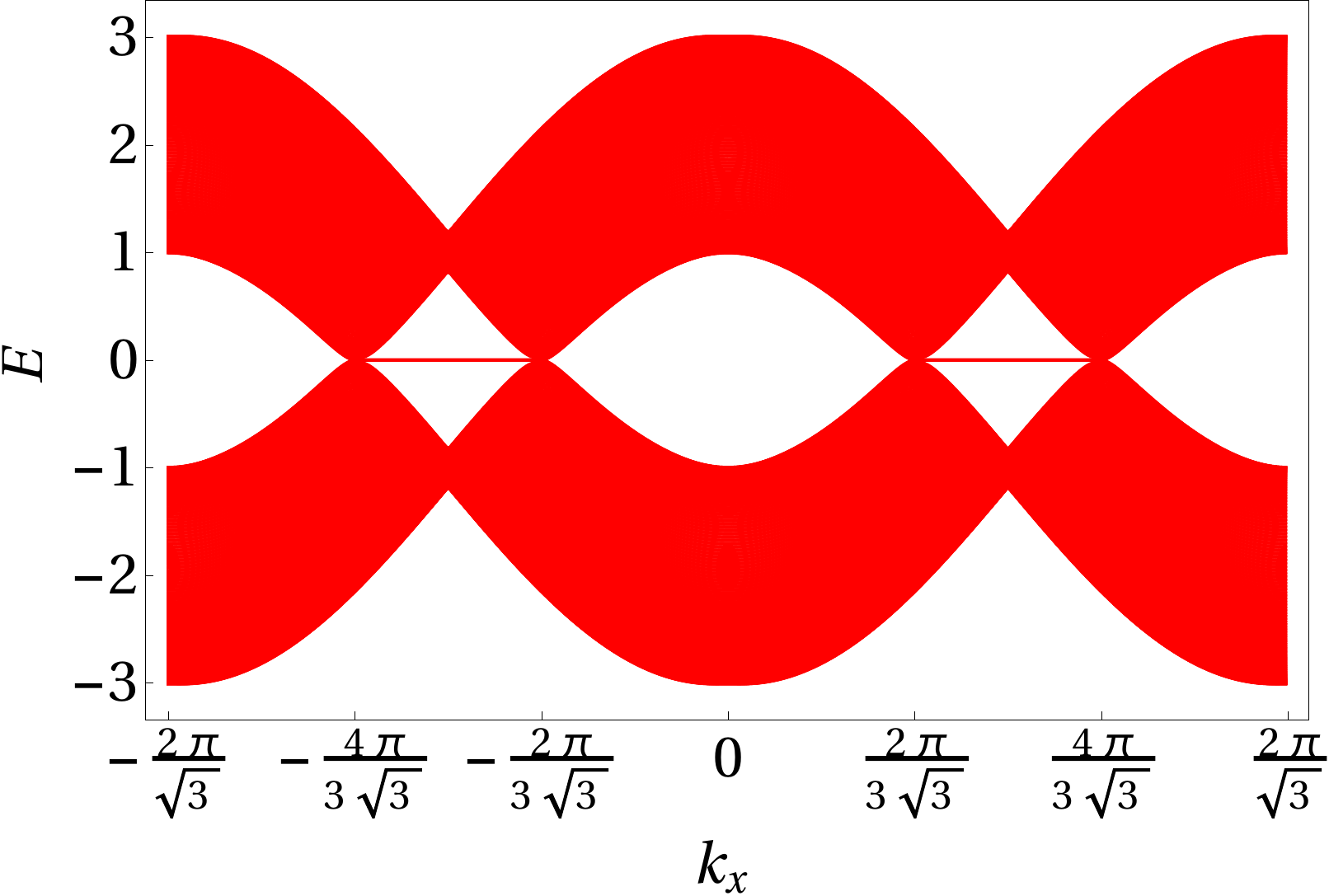}\label{fig:8a}}\hspace{0.2
    cm} \subfloat[]{\includegraphics[trim=0 0 0
      0,clip,width=0.45\textwidth]{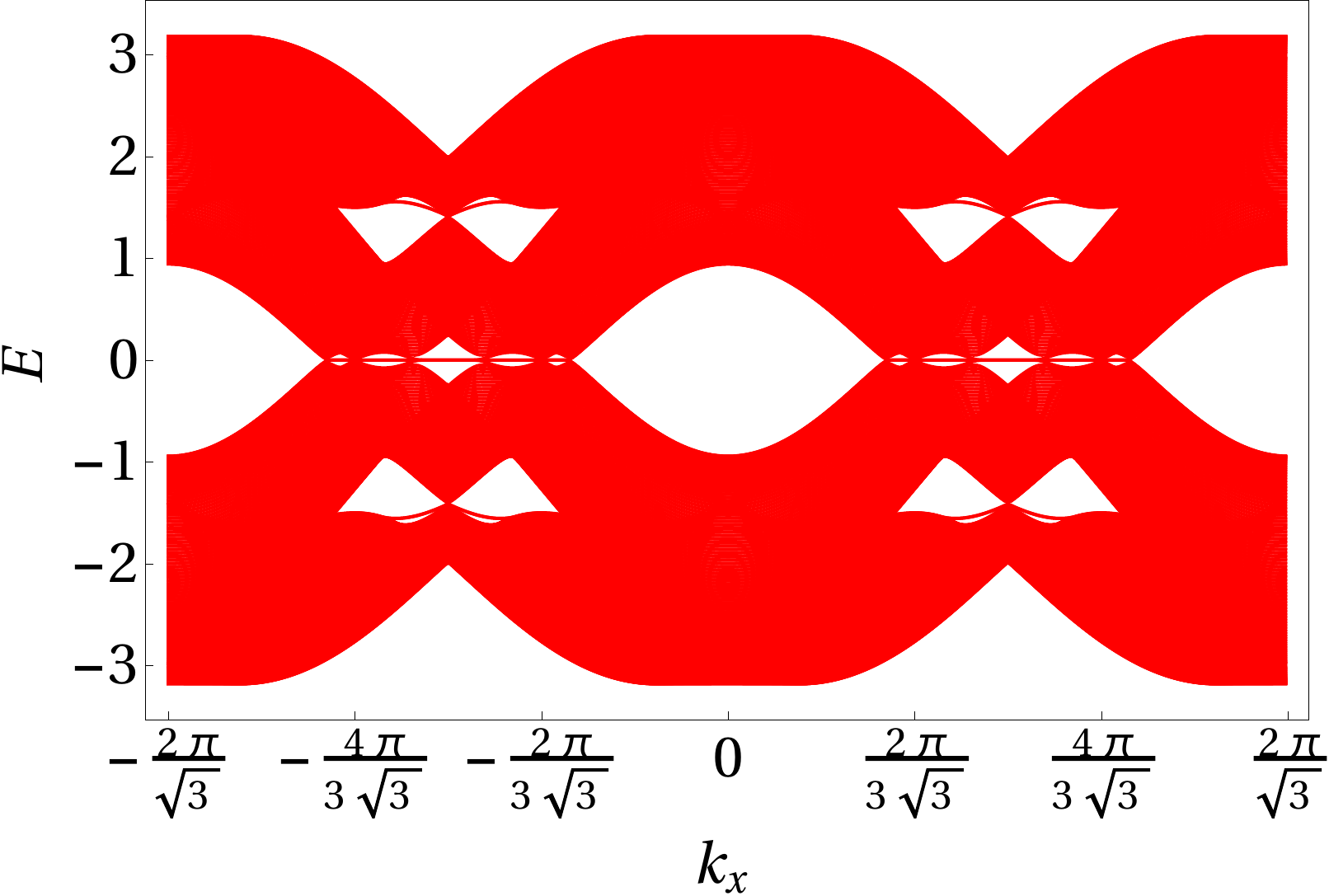}\label{fig:8b}}
  \caption{(color online) Band structure of zigzag ribbon for (a)
    $\lambda_R = 0.1$ and (b) $\lambda_R = 0.5$. Here $t_2=0$.}
\label{rashba1}
\end{figure}

Next we add a Rashba spin orbit coupling to a pristine graphene. Thus
here we have $t_2= 0$, but $\lambda_R \neq 0$. For a small value of
$\lambda_R$, the flat band is observed in Fig.~\ref{rashba1} (a) as we
have seen in pristine graphene. However, if we enhance $\lambda_R$,
the flat band reduces as shown in Fig.~\ref{rashba1} for a large value
of $\lambda_R$, namely $\lambda_R = 0.5$, where the flat bands have
almost vanished and in the $|k_x|$ range of
$\left[\frac{2\pi}{3\sqrt{3}} : \frac{4\pi}{3\sqrt{3}}\right]$.


\begin{figure}[!ht]
\centering \subfloat{\includegraphics[trim=0 0 0
    0,clip,width=0.4\textwidth,height=0.35\textwidth]{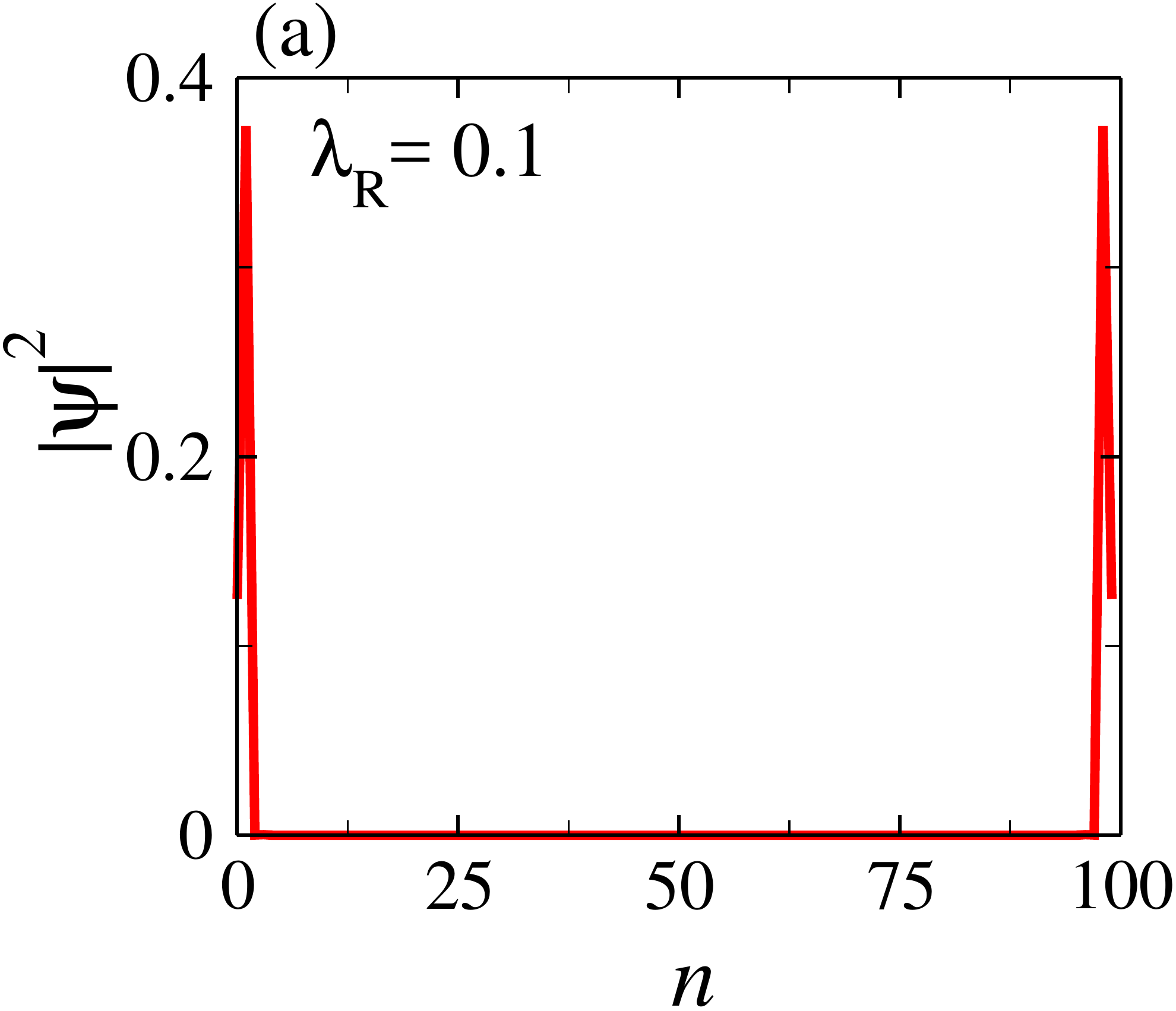}\label{fig:9a}} \hspace{0.2
  cm} \subfloat{\includegraphics[trim=0 -23 0
    0,clip,width=0.4\textwidth,height=0.35\textwidth]{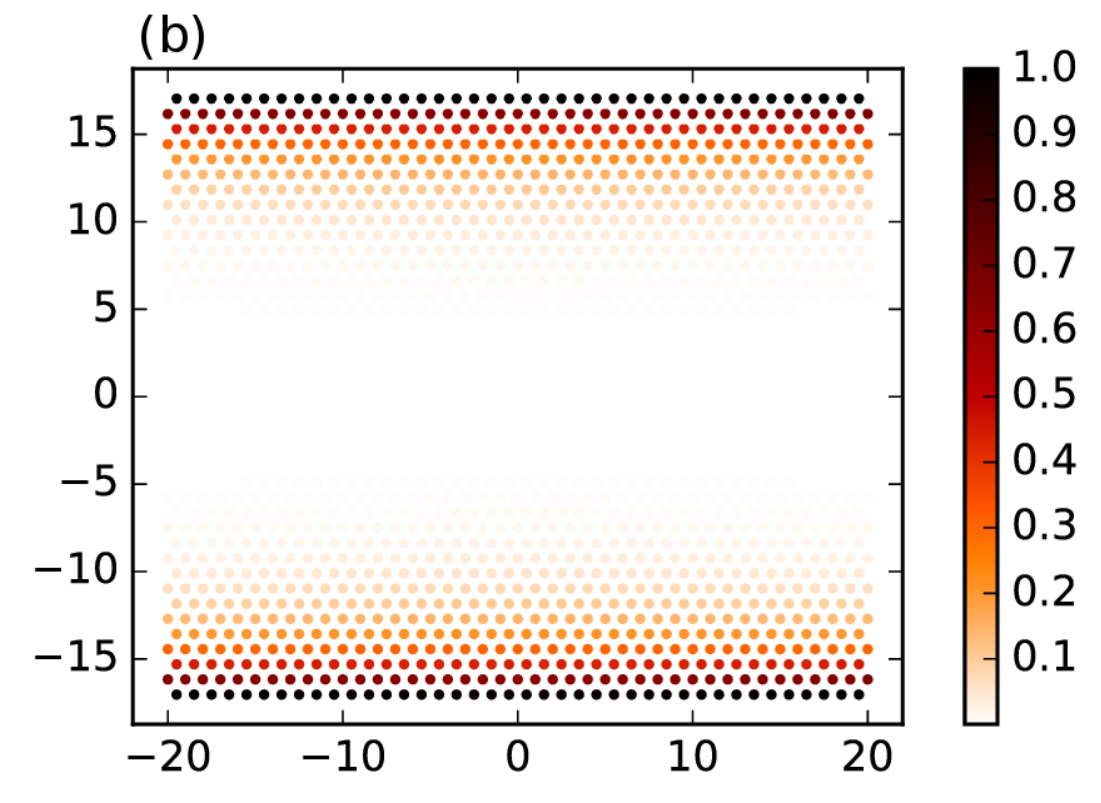}\label{fig:9b}}\\ \subfloat{\includegraphics[trim=0
    0 0
    0,clip,width=0.4\textwidth,height=0.35\textwidth]{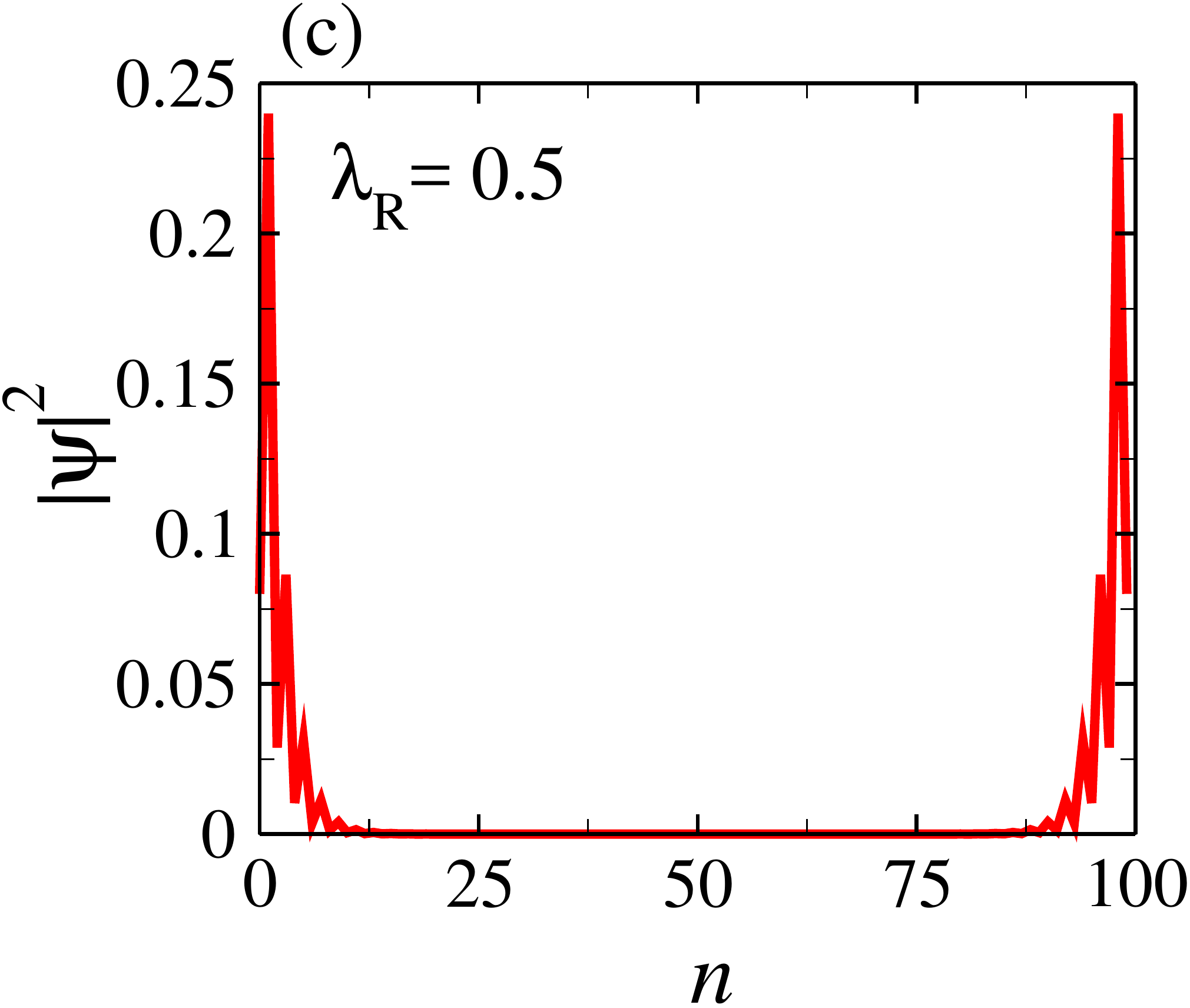}\label{fig:9c}}\hspace{0.2
  cm} \subfloat{\includegraphics[trim=0 -23 0
    0,clip,width=0.4\textwidth,height=0.35\textwidth]{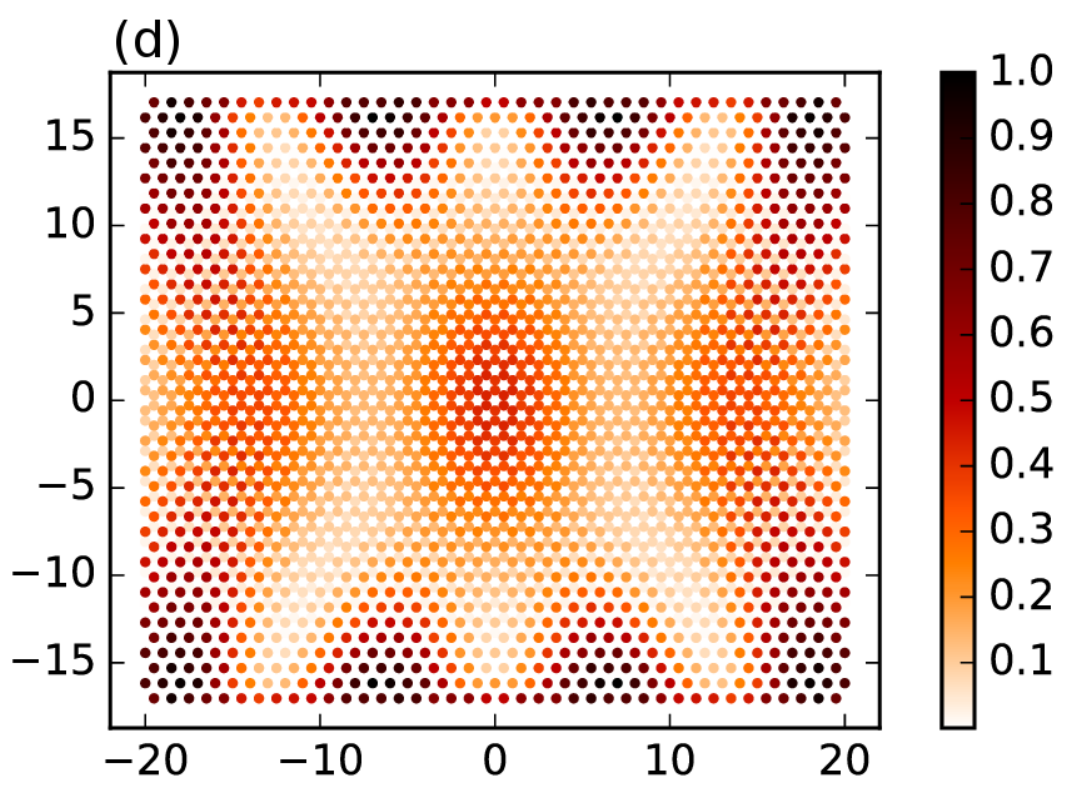}\label{fig:9d}}
  \caption{(Color online) Probability density of the wave-function,
    $|\psi|^{2}$ as a function of $n$ for (a)$\lambda_R
    =0.1$ and (b) $\lambda_R =0.5$. The LDOS plot for (c) $\lambda_R =
    0.1$ (d) $\lambda_R =0.5$. The oscillatory pattern is seen in (c)
    and (d).}
\label{rsoc2}
\end{figure} 

The above features are appropriately justified by the analytic
behavior (obtained by solving Eq. \ref{tripple_eqn2} by putting $t_2
=0$) as shown in Fig.~\ref{rsoc2}. For small values of $\lambda_R$,
there is no oscillation and the probability density decay quickly as
one moves inward as shown in Fig.~\ref{rsoc2} (a). The corresponding
LDOS plot in Fig.~\ref{rsoc2} (b) shows the same behavior. For large
values of $\lambda_R$, the probability densities show damped
oscillations as one moves inside the bulk, and remain finite till
quite a few lattice spacings inside the sample. The LDOS plots in
Fig.~\ref{rsoc2} (d) provides ample support for this oscillatory
behavior and non-vanishing weights inside the bulk.

\begin{figure}[!ht]
  \centering \subfloat{\includegraphics[trim=0 0 0
      0,clip,width=0.4\textwidth,height=0.35\textwidth]{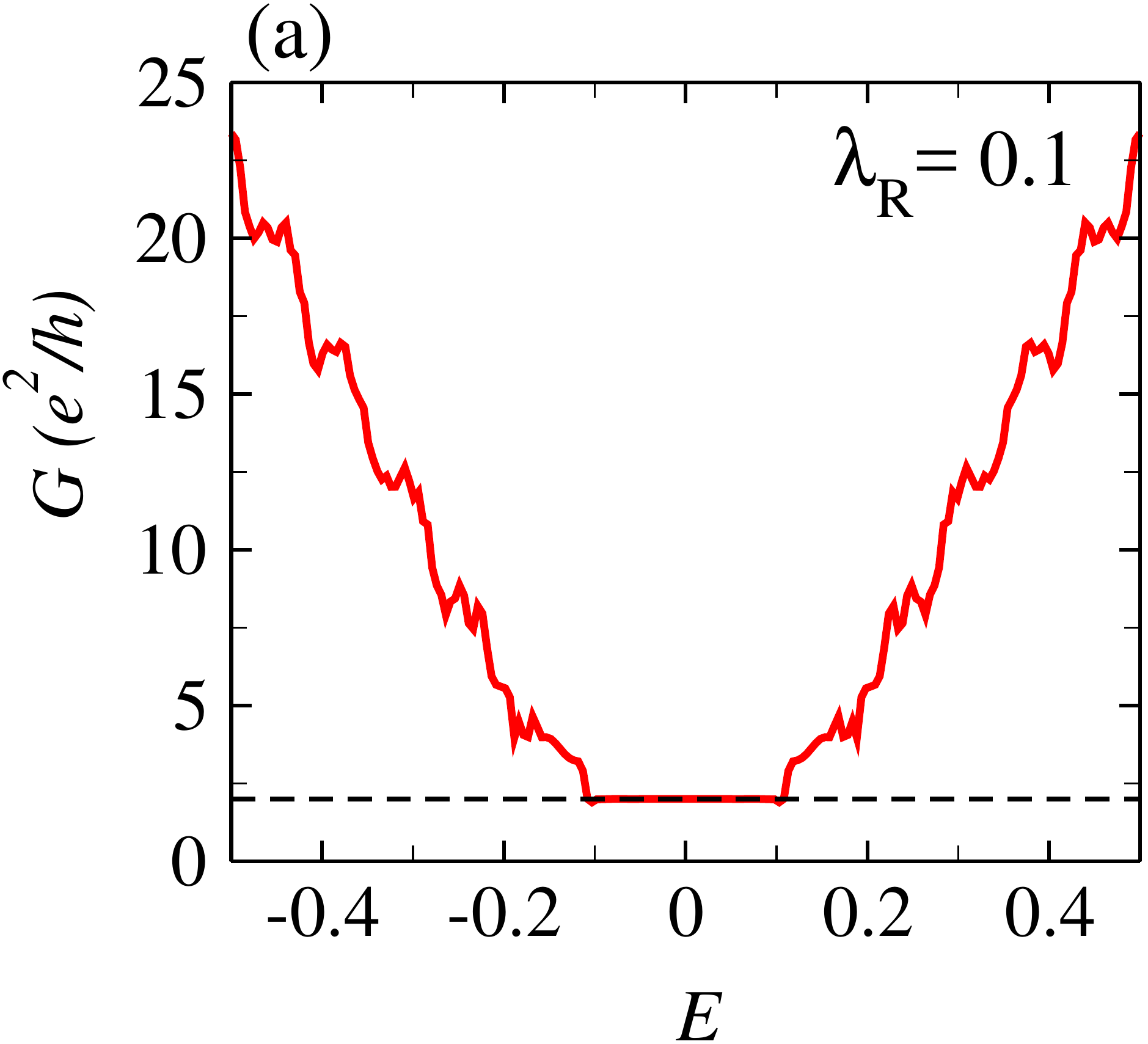}\label{fig:10a}} \hspace{0.2
    cm} \subfloat{\includegraphics[trim=0 0 0
      0,clip,width=0.4\textwidth,height=0.35\textwidth]{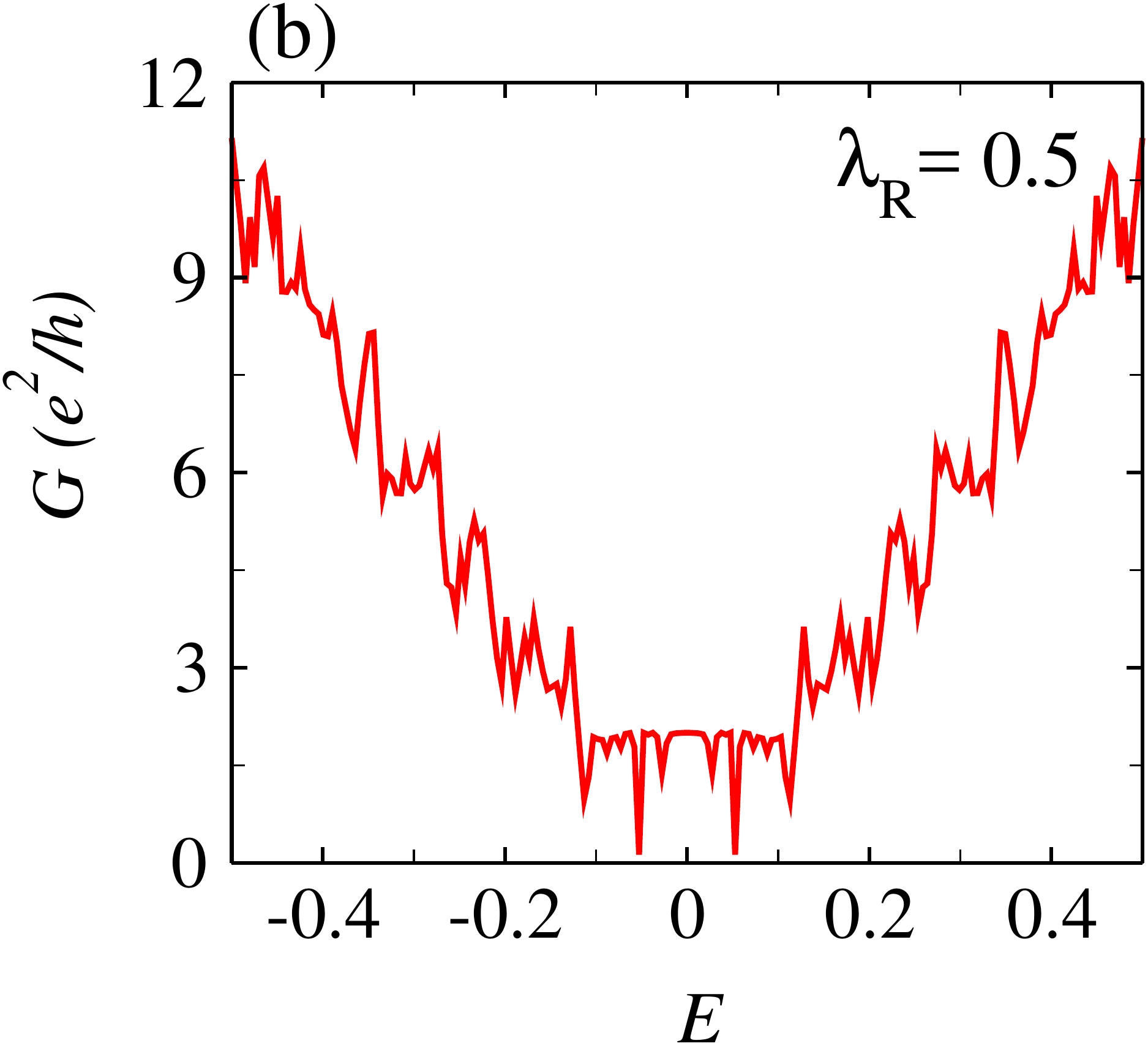}\label{fig:10b}}
  \caption{(color online) The charge conductance, $G$ (in units of
    $e^2/h$) is plotted as a function of energy $E$ (in units of t)
    for (a) $\lambda_R = 0.1$ and (b) $\lambda_R =0.5$}
\label{rsoc3}
\end{figure}

Finally, we have plotted the charge conductance as shown in
Fig.~\ref{rsoc3}. For $\lambda_R=0.1$, there is a $2e^2/h$ plateau
near the zero of the Fermi energy as shown in Fig.~\ref{rsoc3}
(a). However, this $2e^2/h$ value is not associated with a topological
phase as is evident from Fig.~\ref{rsoc2}. The conductance as a
function of energy shows the absence of plateau at a $2e^2/h$ and closing
of gaps are observed near the zero of the Fermi energy in
Fig.~\ref{rsoc3} (b). These results signify the absence of edge modes
and subsequently any topologically non-trivial behavior in the
conductance data.

\subsection{Intrinsic and Rashba SOC}
\label{SOC and RSOC}
In this section, we include both the intrinsic and the Rashba SOC in
the graphene nanoribbon and call this as Kane-Mele nanoribbon (KMNR)
with zigzag edges. It is sensible to ask
what happens to the edge state when both are present. The Kane-Mele Hamiltonian is P-T
symmetric  and the Kramer's doublet must enjoy topological
protection. However, the existence of edge states still needs to be
ascertained and the implications on the conductance spectra thereof.


\begin{figure}[!ht]
  \centering \subfloat[]{\includegraphics[trim=0 0 0
      0,clip,width=0.45\textwidth]{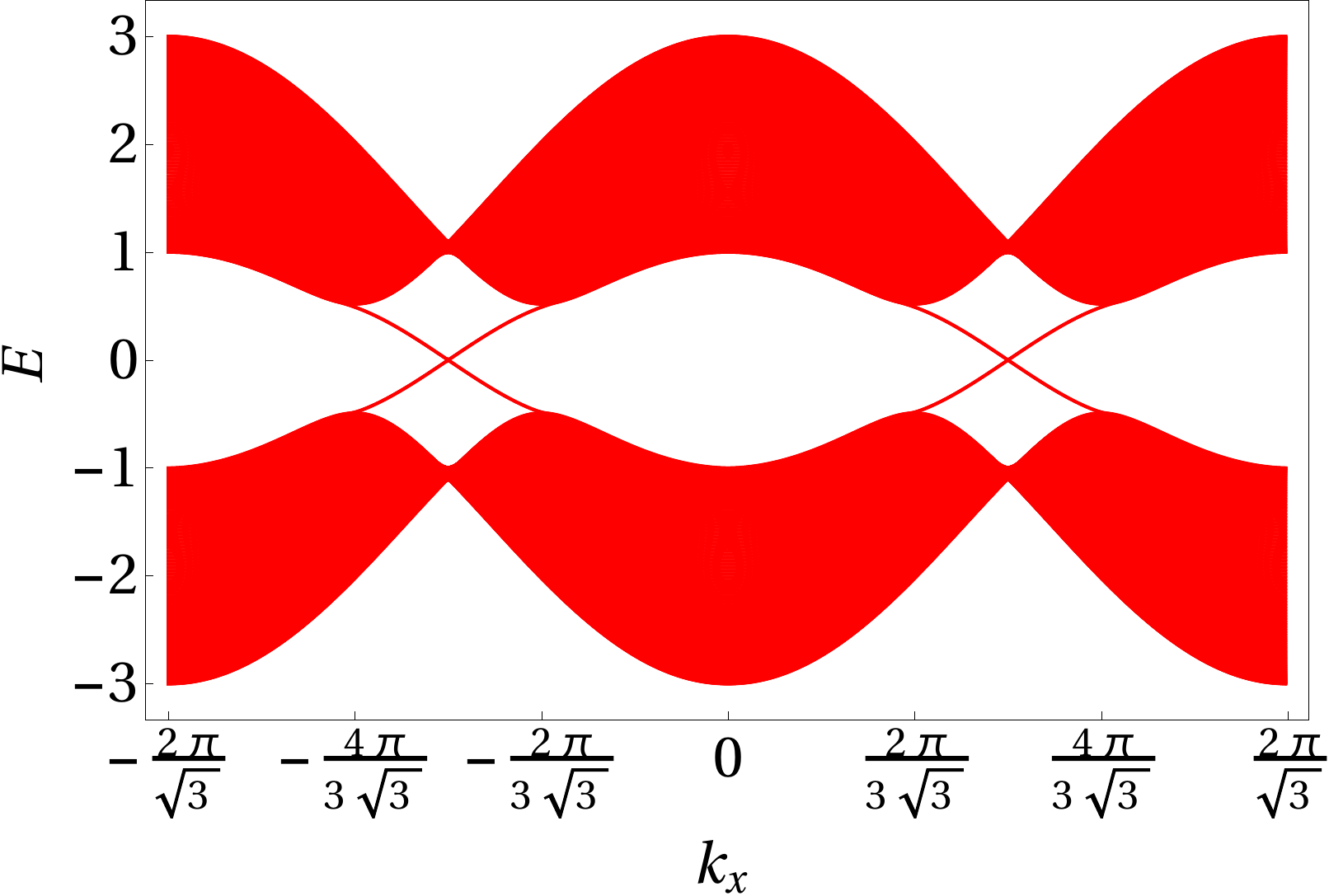}\label{fig:11a}} \hspace{0.2
    cm} \subfloat[]{\includegraphics[trim=0 0 0
      0,clip,width=0.45\textwidth]{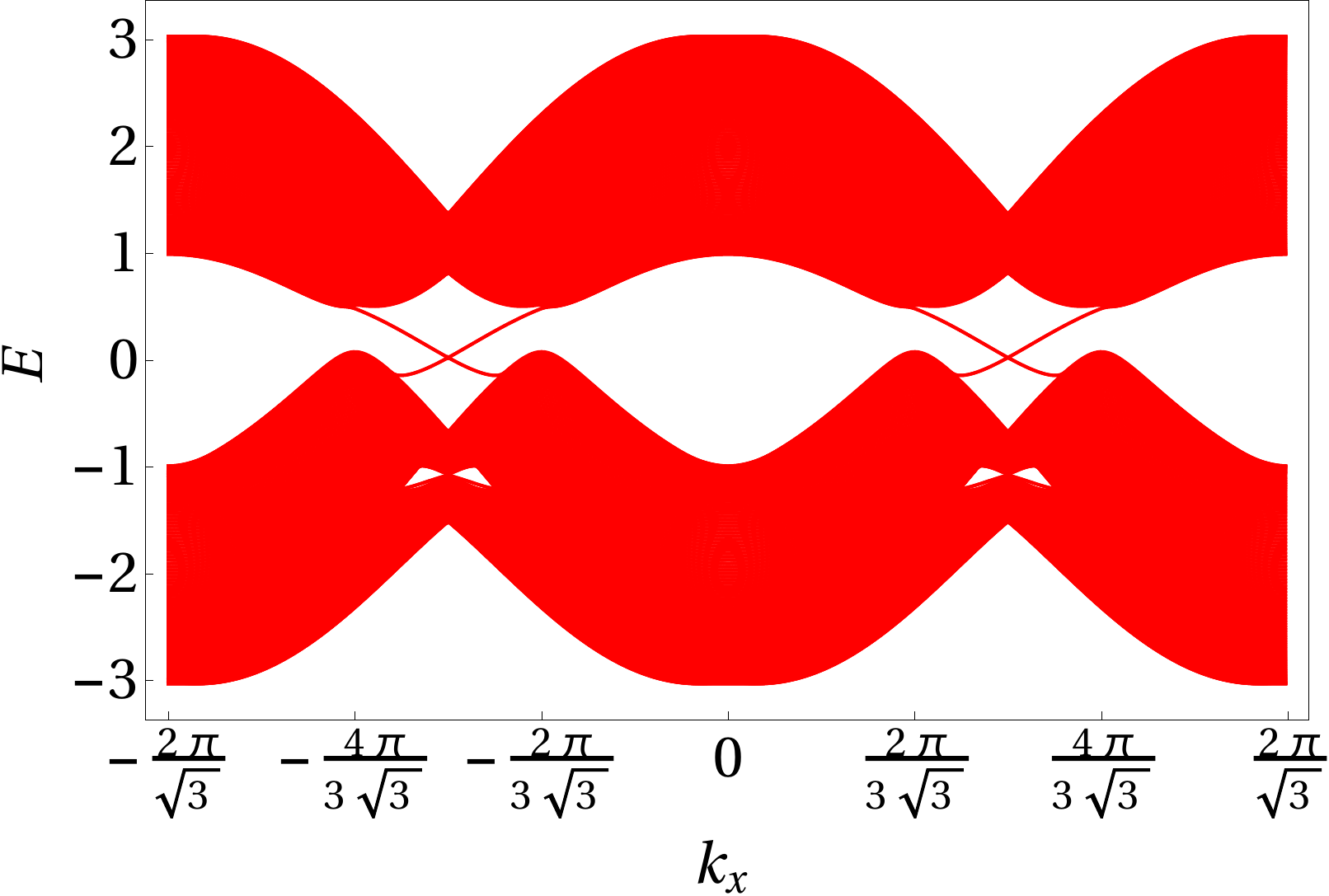}\label{fig:11b}}
  \caption{(color online) The band structure of zigzag KMNR with (a)
    $t_2 = 0.1$ and $\lambda_R = 0.01$ (b) $t_2 = 0.1$ and $\lambda_R
    = 0.2$. }
\label{both1}
\end{figure} 



\begin{figure}[!ht]
  \centering \subfloat{\includegraphics[trim=0 0 0
      0,clip,width=0.4\textwidth,height=0.33\textwidth]{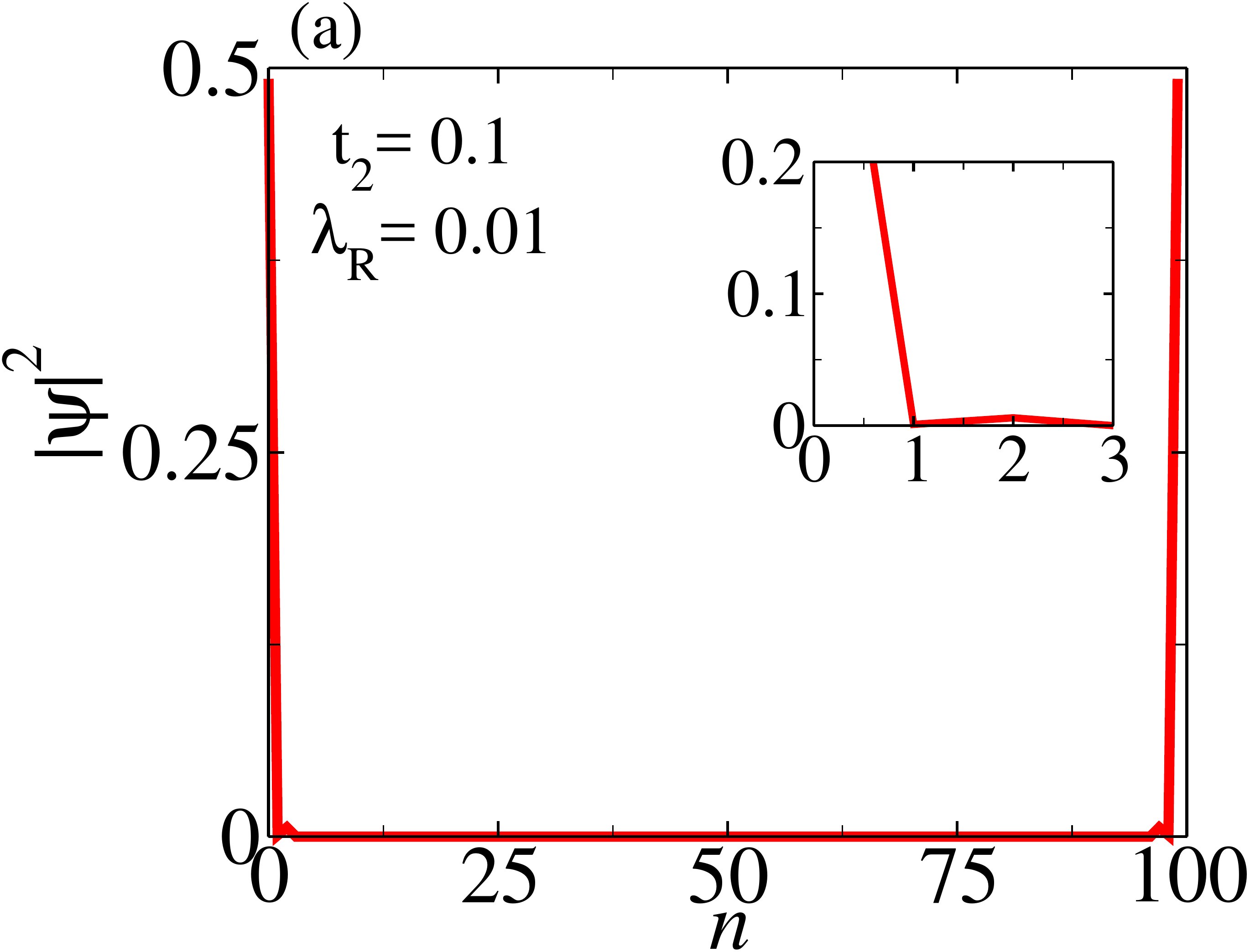}\label{fig:12a}} \hspace{0.2
    cm} \subfloat{\includegraphics[trim=0 -10 0
      0,clip,width=0.4\textwidth,height=0.33\textwidth]{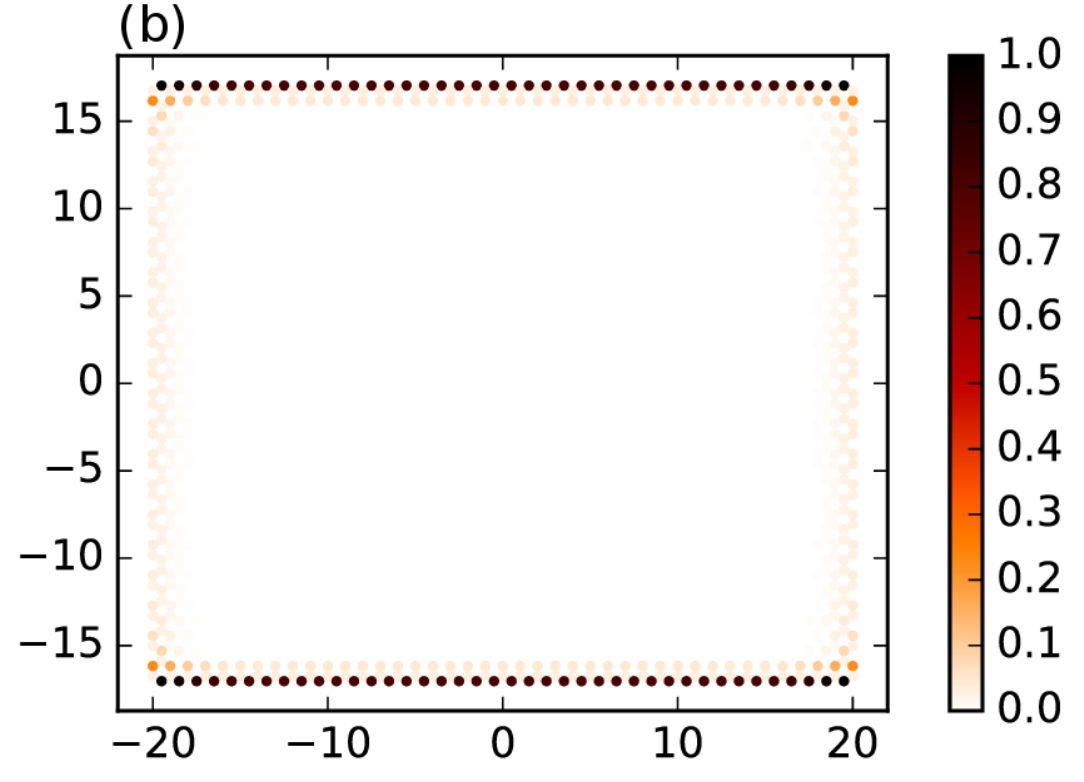}\label{fig:12b}}\\ \subfloat{\includegraphics[trim=0
      0 0
      0,clip,width=0.4\textwidth,height=0.33\textwidth]{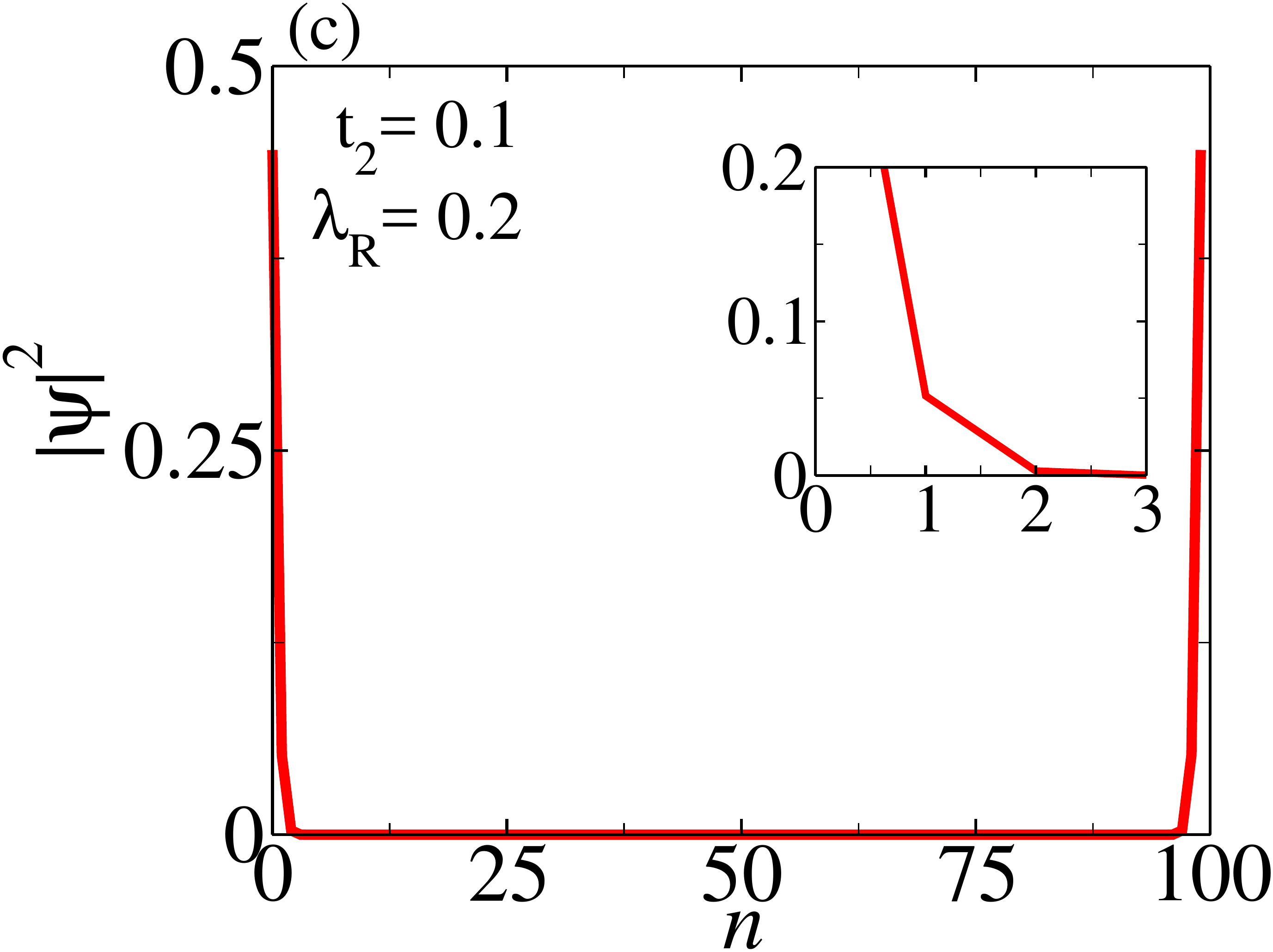}\label{fig:12c}} \hspace{0.2
    cm} \subfloat{\includegraphics[trim=0 -10 0
      0,clip,width=0.4\textwidth,height=0.33\textwidth]{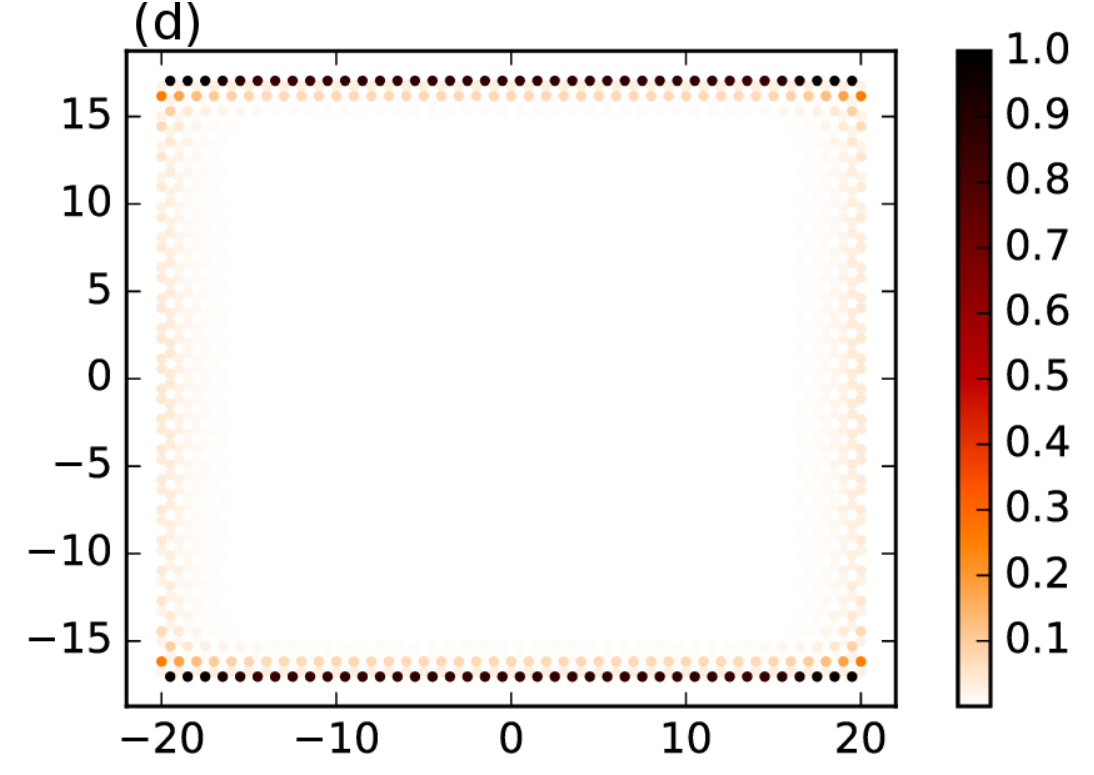}\label{fig:12d}}
  \caption{(color online) (a) Probability density of the
    wave-function, $|\psi|^{2}$ as a function of $n$ and
    (b) the corresponding LDOS plot for $t_2 = 0.1$,$\lambda_R =
    0.01$. (c) and (d) is plotted for $t_2 = 0.1$ and $\lambda =
    0.2$.}
\label{both2}
\end{figure}



  We keep $t_2 =0.1$ and explore two values of Rashba SOC, namely, $\lambda_R = 0.01$ and $\lambda_R = 0.2$. The former corresponds to $\lambda_R < t_2$ $(<t)$ and the latter denote $\lambda_R > t_2$ $(<t)$. The band structures for the corresponding cases are shown in Fig. \ref{both1}. When the strength of the Rashba SOC is weak, the band structure for $t_2= 0.1$ is almost similar (shown in Fig. \ref{both1} (a)) with the band structure for the same value of the intrinsic SOC in the absence of Rashba coupling (Fig. \ref{soc1} (a)). However, for a higher value of $\lambda_R$ (Fig. \ref{both1} (b)), the band gap gets smaller than the previous case. The results corresponding to the two cases are not much different with regard to the existence of the edge states. The only (minor) difference is that the analytic form yields a non-zero value for $n=2$ corresponding to the larger value of $\lambda_R$, that is, $\lambda_R= 0.2$. The LDOS maps corroborate existence of edge modes (see Fig. \ref{both2} (b) and (d)) as is evident for Fig. \ref{both2} (a) and (c).


\begin{figure}[!ht]
  \centering \subfloat{\includegraphics[trim=0 0 0
      0,clip,width=0.4\textwidth,height=0.35\textwidth]{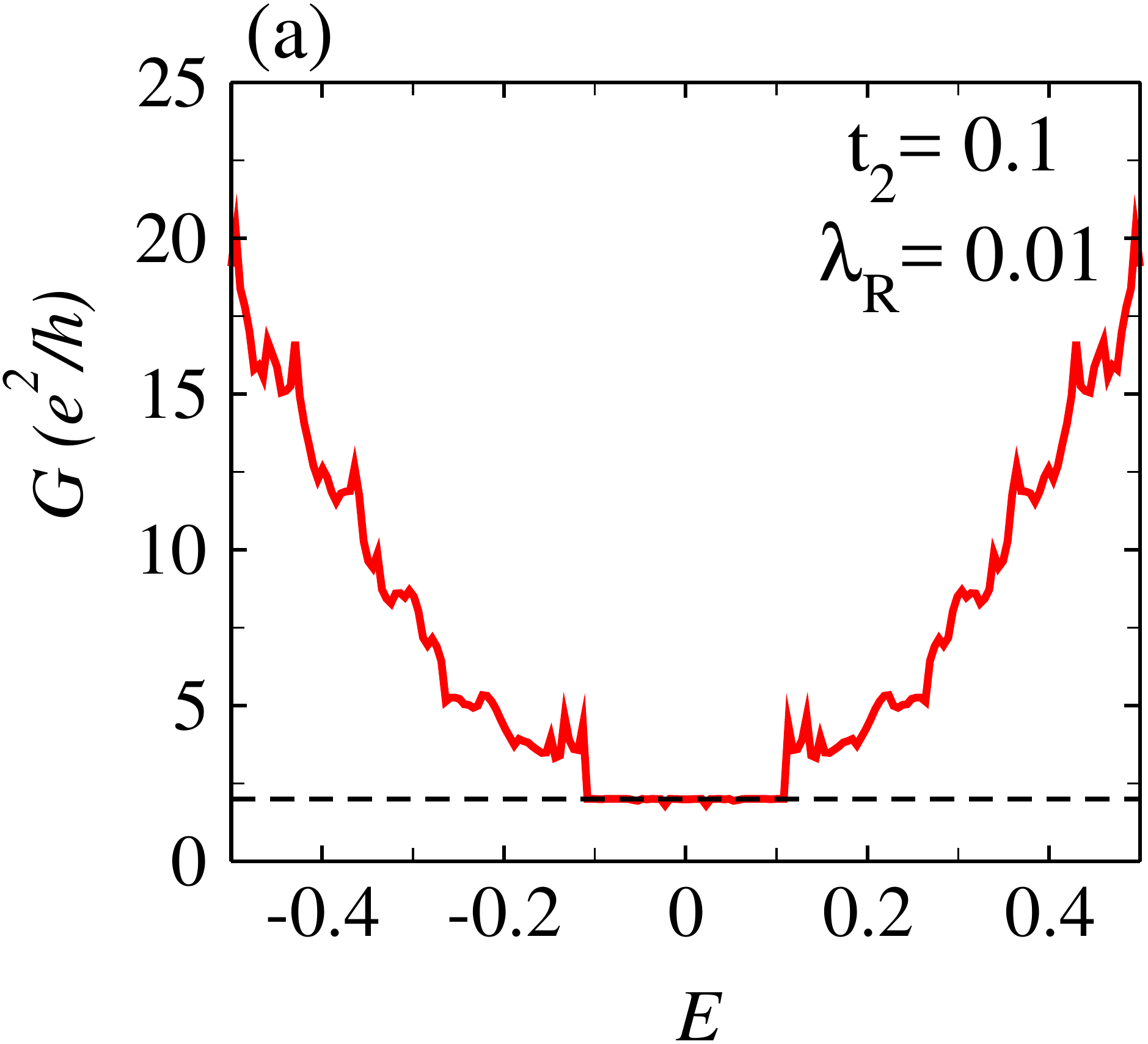}\label{fig:13a}} \hspace{0.2
    cm} \subfloat{\includegraphics[trim=0 0 0
      0,clip,width=0.4\textwidth,height=0.35\textwidth]{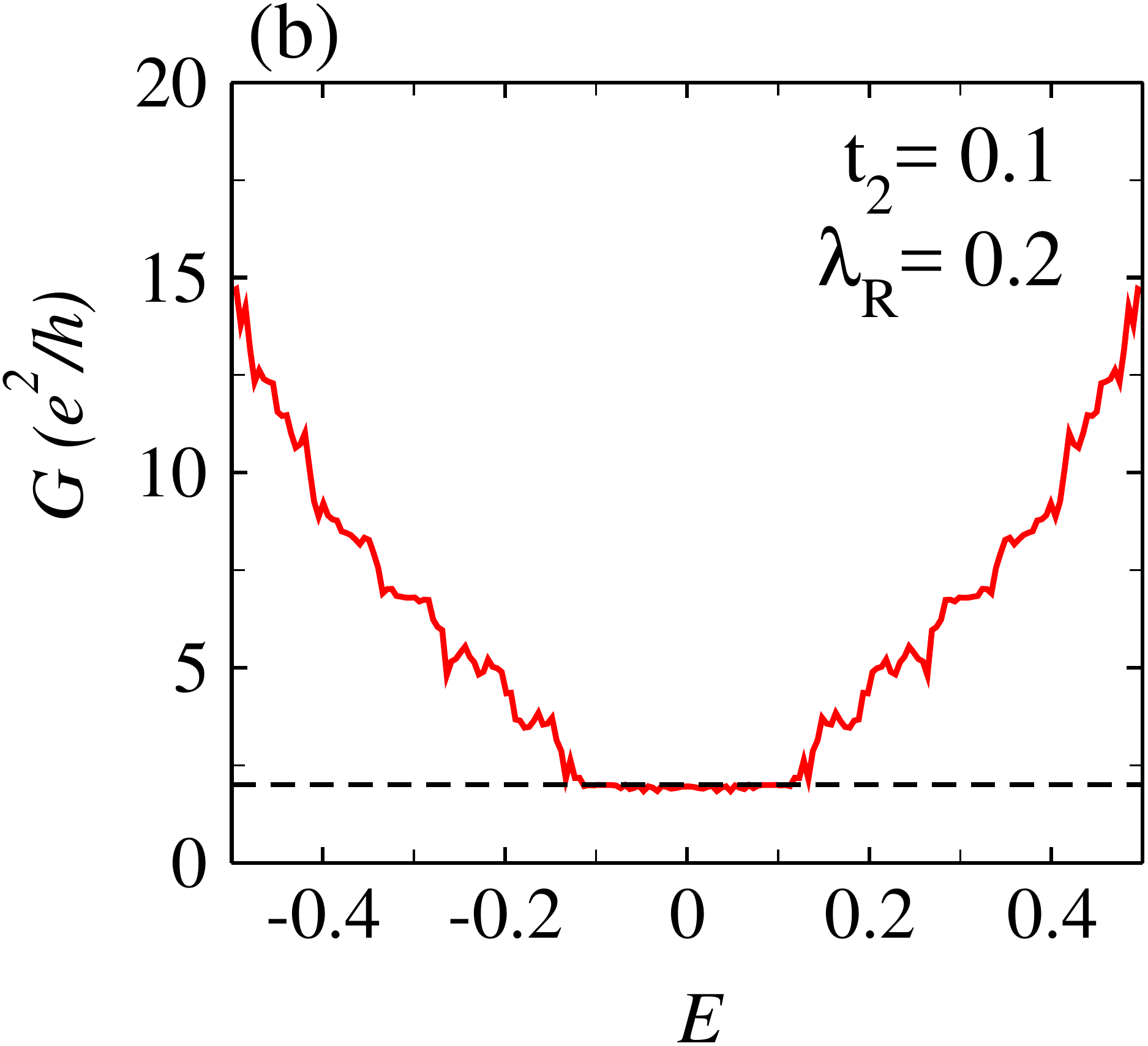}\label{fig:13b}}
  \caption{(color online) The charge conductance, $G$ (in units of
    $e^2/h$) is plotted as a function of energy $E$ (in units of t)
    for $t_2 =0.1$ and $\lambda_R = 0.2$.}
\label{both3}
\end{figure}


The conductance plots as a function of the Fermi energy in
Fig.~\ref{both3} show the existence of a $2e^2/h$ plateau, which are topologically protected and correspond to quantum spin Hall
insulating phase. It may be noted that we have presented plots only
for two sets of parameter values, namely ($t_2$, $\lambda_R$) = (0.1,
0.01) and (0.1, 0.2). However, the same inferences can be drawn for
other sets, such as larger values of $t_2$ and $\lambda_R$ which we
have explicitly checked, but have not shown the data for brevity.


\section*{Conclusions}
In this paper, we have analytically computed expressions for the edge
modes for a Kane-Mele nanoribbon (KMNR). The analytic results are
supported by LDOS obtained using KWANT. Further, we have calculated the
band structure and conductance for a pristine graphene, graphene with intrinsic SOC, graphene with Rashba SOC and graphene with both SOCs. We re-establish the existence of
topologically protected edge states owing to the presence of parity and
time reversal symmetry of the Hamiltonian. The system acquires edge states in presence of spin orbit couplings as observed
from band structure and both analytic and numeric calculation of
electron probability densities. The conductance spectra further show
a plateau at a non-zero value $(=2e^2/h)$ near the zero of the Fermi
energy.

\setcounter{secnumdepth}{0}
\section{ACKNOWLEDGMENTS} 
SB thanks SERB, India for financial support under the grant F. No: EMR/ 2015/001039.


\begin{thebibliography}{99}
\bibitem{Priyanka} $n$ is a number such that $k_{x}$ is bounded between $\left[\frac{2\pi}{3\sqrt{3}} : \frac{4\pi}{3\sqrt{3}}\right]$. Here we have taken $n$ = 2.15.
\end{thebibliography}
\end{document}